\pdfoutput=1
\documentclass[journal,twoside,web]{ieeecolor}
\usepackage{tmi}
\usepackage{cite}
\usepackage{xcolor}  %

\usepackage{format}

\newcommand{\ObsOp}{\bE}

\newcommand{\pisim}{\mathbf{sim}_{\pi}}
\newcommand{\pidotsim}{\mathbf{dotsim}_{\pi}}
\newcommand{\pidistsim}{\mathbf{distsim}_{\pi}}
\newcommand{\bb}{\bm{b}}
\newcommand{\bvarphi}{\boldsymbol{\varphi}}
\newcommand{\bdelta}{\boldsymbol{\delta}}

\newcommand{\longi}[2]{{#1}_{(#2)}}

\newcommand{\longij}[3]{{#1}^{#2}_{#3}}
\newcommand{\GGT}{\mathrm{GGT}}

\newcommand{\note}[1]{}

\def\BibTeX{{\rm B\kern-.05em{\sc i\kern-.025em b}\kern-.08em
    T\kern-.1667em\lower.7ex\hbox{E}\kern-.125emX}}
\begin{document}
\title{LGGS: Longitudinal MRI Reconstruction via Guided Group Sparsity}

\author{%
Nikola Janju\v{s}evi\'{c},~\IEEEmembership{Member,~IEEE},
Hersh Chandarana,
Yao Wang,~\IEEEmembership{Fellow,~IEEE},
and Li Feng
\thanks{This work was supported in part by National Institute of Biomedical
Imaging and Bioengineering (NIBIB) grants R01~EB030549, R01~EB031083, and
P41~EB017183. The authors thank NYU Langone Health UltraViolet for technical
support, and Dan Sodickson and Arda Atal{\i}k for their helpful discussions.}
\thanks{N. Janju\v{s}evi\'{c}, H. Chandarana, and Li Feng are with the Radiology Department,
New York University Grossman School of Medicine, New York, NY 10016 USA
(e-mail: nikola.janjusevic@nyulangone.org; hersh.chandarana@nyulangone.org; li.feng@nyulangone.org).}
\thanks{Y. Wang is with the Electrical and Computer Engineering Department,
New York University Tandon School of Engineering, Brooklyn, NY 11201 USA
(e-mail: yw523@nyu.edu).}
\vspace*{-20pt}}

\maketitle

\begin{abstract}
    Serial MRI is routinely performed in current clinical practice for disease
    surveillance and applications such as MR-guided radiotherapy. However, long
    MRI scan times increase costs and limit patient access, making accelerated
    image reconstruction of broad clinical importance. Previous scans of the
    same patient contain image information highly correlated with the current
    exam, yet existing reconstruction methods rarely exploit this information.
    Major challenges in the effective use of prior scans include inter-visit
    misregistration, pathological changes, and image phase variation. In this
    work, we propose Longitudinal Guided Group Sparsity (LGGS), an
    interpretable deep-learning framework for longitudinally informed MRI
    reconstruction derived from classical group sparsity. The prior scan is
    incorporated into the reconstruction regularizer through learned
    phase-invariant circulant attention that identifies corresponding features
    across exams. This formulation provides robustness to local in-plane and
    through-plane misregistration as well as inter-visit phase variation.
    Across acceleration rates, LGGS achieves state-of-the-art performance on
    longitudinal T1-weighted multi-coil brain MRI data, outperforming the
    evaluated unguided and longitudinally informed reconstruction methods, with
    substantially fewer learned parameters. We further
    evaluated the contributions of multiple guide slices and phase-invariant
    similarity and demonstrated that accelerated prior images can also serve as
    effective guides with minimal degradation in reconstruction quality.
\end{abstract}

\bstctlcite{IEEEexample:BSTcontrol}

\section{Introduction and Background}\label{sec:intro}
\begin{figure}[tb]
    \centering
    \includegraphics[width=\columnwidth]{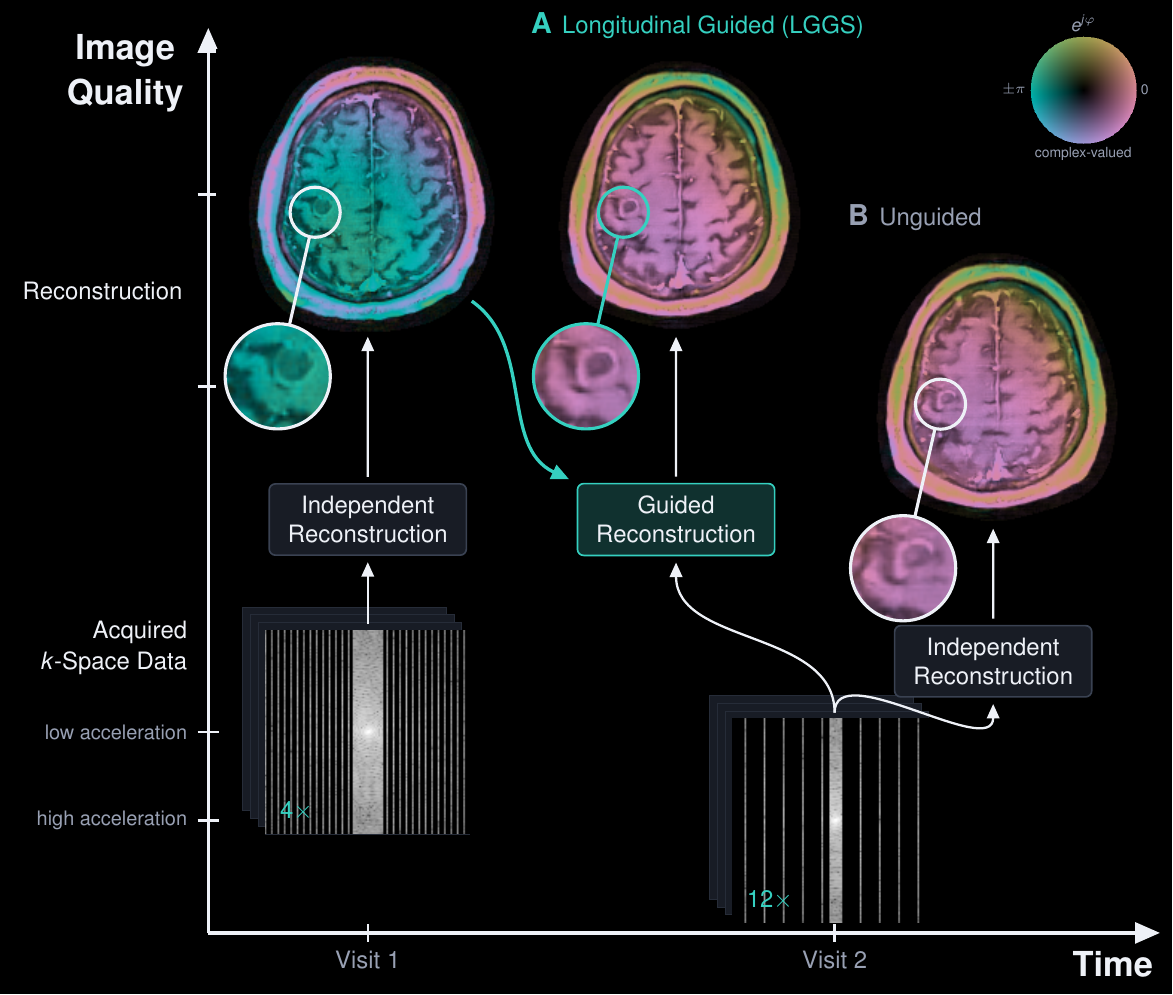}
    \caption{%
        Longitudinal accelerated multi-coil MRI reconstruction. 
        Images are shown as complex-valued, hue encoding phase (key, top right). 
        \textbf{(A)} LGGS takes a prior visit acquired at $4\times$ as a
        \emph{guide} for a current visit acquired at $12\times$. \textbf{(B)}
        An unguided reconstruction of the same measurements discards it and
        fails to resolve pathology development (see zoom region). Patient
        image data is non-trivially related over visits due to image
        misregistration (in-plane and through-plane), disease progression, and
        inter-visit phase change. The issue of inter-visit phase remains even
        for methods using magnitude-based guidance as relative phase difference
        to the complex-valued reconstruction persists.
    }
    \label{fig-framework}
\end{figure}

\IEEEPARstart{S}{erial} magnetic resonance imaging (MRI), in which patients
undergo repeated exams over days, months, or years, is an essential component of
clinical practice for disease surveillance, treatment response
assessment, and applications such as MR-guided
radiotherapy~\cite{wattjes2021magnims, wen2023rano, hall2019mrlinac}. However,
MRI exams can be time-consuming, with high-resolution, multi-sequence protocols
often requiring tens of minutes to an hour. These long exam times limit clinical
throughput and patient tolerance~\cite{fastmri}, particularly for patients who
require frequent follow-ups.

MRI acquisition can be accelerated by reducing the amount of acquired k-space
data and using advanced reconstruction methods to recover images while
suppressing the resulting undersampling artifacts. Classical methods such as
sensitivity encoding (SENSE)~\cite{pruessmann1999sense} and compressed sensing
(CS)~\cite{lustig2007sparse} exploit multi-coil sensitivity information and
transform-domain sparsity, respectively. More recently, deep-learning
reconstruction methods~\cite{ongie2020deep, sriram2020end, lpds2025} have
demonstrated substantial improvements in reconstruction quality by exploiting
image priors learned from large datasets, and some have been adopted for
clinical use~\cite{siemans-deep-resolve}. However, most existing reconstruction
methods treat each exam independently and do not exploit prior MR images often
available from the same patient. These prior exams provide highly correlated,
patient-specific anatomical information at no additional acquisition cost,
offering a valuable opportunity to further accelerate MRI reconstruction.
However, incorporating prior images into the reconstruction of the current exam
is not straightforward, as this information must be used carefully to preserve
clinically important changes between exams rather than bias the reconstruction
towards outdated anatomy (Figure~\ref{fig-framework}).

The use of prior images to improve reconstruction has been explored in several
studies. In computed tomography (CT), prior-image constrained compressed
sensing (PICCS)~\cite{chen2008piccs} penalized the deviation of the
reconstruction from an earlier image. More recently, several methods have
incorporated previous MRI exams into MRI reconstruction. Chen \emph{et
al.}~\cite{Chen2026} concatenated prior visits to improve dynamic MRI
reconstruction via a low-rank subspace formulation. Urman \emph{et
al.}~\cite{urman2024longitudinal} used prior images to initialize
diffusion-based posterior reconstruction, while Shamaei \emph{et
al.}~\cite{shamaei2025prior} combined an initial reconstruction with prior-image
registration and transformer-based refinement. The longitudinal Trust-Guided
Variational Network (L-TGVN)~\cite{ltgvn2026}, an extension of
TGVN~\cite{tgvn2026}, incorporated prior exams through a learned guidance
network conditioned on both the prior images and the current reconstruction to
improve image quality. However, effective use of longitudinal information
presents several important challenges:
\begin{itemize}
    \item \textbf{Misregistration.} Differences in patient positioning between
    visits can introduce in-plane misregistration (translation and rotation)
    and out-of-plane misregistration (slice misalignment). Global alignment can
    be achieved through integer-voxel registration using low-resolution images
    reconstructed from the fully sampled central k-space region, which is
    commonly available in MRI acquisitions. However, residual local
    misregistration may persist, making direct image-difference comparisons
    between prior and current images unreliable and potentially detrimental to
    reconstruction quality.

    \item \textbf{Pathological change.} Disease progression or treatment
    response may alter anatomical structure between visits. A na\"{i}ve
    reconstruction method which assumes the guide closely matches the target
    can be misled by such changes, potentially masking clinically relevant
    findings.

    \item \textbf{Image phase variation.} The phase of a reconstructed MR image
    is influenced by magnetic field inhomogeneity, coil loading, and coil
    sensitivity estimation~\cite{robinson2017phase}. These factors can vary
    between exams and introduce phase differences between the prior and current
    images. Such phase differences generally do not reflect anatomical changes
    but can compromise conventional feature similarity measures and direct
    image-difference comparisons.
\end{itemize}

In this work, we propose \textbf{Longitudinal Guided Group Sparsity (LGGS)}, an
interpretable deep-learning framework that incorporates information from prior
MRI exams directly into the regularizer of an unrolled reconstruction
network~\cite{monga2021unrolling} while addressing all of the challenges
outlined above. LGGS builds upon two approaches from our prior work: (1) the
analysis-domain sparse-coding framework of LPDSNet~\cite{lpds2025} and (2) the
learned group-sparsity formulation of GroupCDL~\cite{janjusevic2025groupcdl},
both developed for MRI reconstruction. Rather than directly incorporating
prior-image features into the current reconstruction, LGGS uses their
correspondence with current features to adapt reconstruction regularization. A
novel phase-invariant circulant attention mechanism is developed to identify
corresponding features across exams despite residual spatial misregistration and
inter-visit phase differences.

The main contributions are summarized as follows:
\begin{itemize}
    \item \textbf{Guided group-sparsity regularization}: We introduce a novel
        regularizer that uses prior-image features to adapt the regularization
        of the current reconstruction through learned, row-normalized feature
        associations. The group associations are constructed only between the
        current reconstruction and prior exams, without direct transfer of
        prior-image features. In the absence of prior information, the
        regularizer reduces to the standard $\ell_1$ penalty used in the
        unguided LPDSNet~\cite{lpds2025}.

    \item \textbf{Phase-invariant circulant attention}: We develop a novel
        phase-invariant similarity measure to identify corresponding features
        between current and prior exams despite inter-visit phase differences.
        A local search window and neighboring guide slices further improve
        robustness to residual spatial misregistration, while feature
        similarity determines the influence of prior information on
        reconstruction regularization.

    \item \textbf{Parameter-efficient longitudinal reconstruction}: We develop
        an interpretable longitudinal reconstruction network that achieves
        superior reconstruction quality compared with the evaluated unguided
        and longitudinally informed methods across different acceleration
        rates, with substantially fewer learned parameters. Our proposed method
        also shows effective reconstruction when the prior images are
        themselves accelerated reconstructions.
\end{itemize}

\section{Preliminaries and Related Work} \label{sec:prelim}
\begin{table}[tb]
\centering
\caption{Notation. \label{tab:notation}}
\vspace{0.5em}
\resizebox{\linewidth}{!}{%
\begin{tabular}{|l|l|} \hline
\multirow{2}{*}{$\x \in \C^{NC}$} & a vector-valued image with $N=N_1\times N_2$ pixels, 
\\ & and vectorized channels, $\x =[ \x_1^\top,\, \cdots,\, \x_C^\top ]^\top$. \\
\hline
$\x_c \in \C^N$ & the $c$-th channel/coil of $\x$. \\
\hline
$\x[n] \in \C^C$ & the $n$-th pixel of $\x$, $n \in [1, \, N]$. \\
\hline
\multirow{2}{*}{$\bU \otimes \bV \in \C^{QM \times NC}$} & Kronecker product of $\bU \in
\C^{Q\times N}$ and $\bV \in \C^{M\times C}$, \\
& i.e. the block matrix with $\bU$s scaled by $\bV_{ij} ~ \forall ~ i,j$.\\
\hline
\multirow{2}{*}{$\y = \cwise{\bU}\x \equiv (\bU \otimes \IDMAT_C)\x$} & the matrix $\bU$ applied channel-wise, \\
& i.e. $\y_c = \bU\x_c \in \C^Q,~\forall~1 \leq c \leq C$. \\
\hline
\multirow{2}{*}{$\y = \pwise{\bV}\x \equiv (\IDMAT_N \otimes \bV)\x$} & the matrix $\bV$ applied pixel-wise,\\
& i.e. $\y[n] = \bV\x[n] \in \C^M,~\forall~1 \leq n \leq N$. \\
\hline
\end{tabular}}
\end{table}

\subsection{Accelerated and Longitudinal MRI}
We consider the MR acquisition of subsampled multi-coil k-space data
${\y \in \C^{N_s C}}$ from ground-truth image ${\x \in \C^N}$ as,
\begin{equation}\label{eq-forward}
\y_c = \IDMAT_{\Omega}\bF( \bs_c \odot \x ) + \bnu_c,
\quad \forall~c=1,\dots,C,
\end{equation}
where ${\bs_c \in \C^{N}}$ denotes the $c$-th coil sensitivity map,
${\odot}$~denotes element-wise multiplication, $\bF$ denotes the $N$-pixel
2D DFT matrix, $\IDMAT_\Omega$ denotes the row-removed identity matrix with
kept rows indicated by index set $\Omega$, and
${\bnu \in \C^{N_s C}}$ denotes pixel-wise i.i.d.\ complex Gaussian noise
with covariance $\bSigma_y$, i.e.\ ${\bnu[n] \sim \CN(0,\,\bSigma_y)}$
for all ${n \in [1, N_s]}$.
Following the notation of~\cite{Rep2Rep2025}, \eqref{eq-forward} may be
written as the single vector equation ${\y = \ObsOp\x + \bnu}$, where
${\ObsOp = \cwise{\IDMAT_{\Omega}\bF}\bS}$ and
${\bS = \begin{bmatrix} \diag(\bs_1) & \cdots & \diag(\bs_C) \end{bmatrix}^\top
\in \C^{NC \times N}}$. See Table~\ref{tab:notation} for further notational
details.

\subsection{Guided Reconstruction}

Several recent works incorporate a prior examination of the same patient into
MRI reconstruction. Urman \emph{et al.}~\cite{urman2024longitudinal} propose a
latent diffusion framework in which the prior examination steers posterior
sampling at inference, requiring no longitudinally paired training data.
However, the guide is admitted only through the initialization of the sampling
trajectory of a generative prior, and so enforces no structural correspondence
between the prior scan and the reconstruction.

The Trust-Guided Variational Network (TGVN)~\cite{tgvn2026} presents \soa
results on accelerated MRI reconstruction with image side information serving as
a ``guide,'' and its longitudinal variant, L-TGVN~\cite{ltgvn2026}, is the \soa
for the prior-scan setting considered in this work. Both build upon the End-to-End
Variational Network (E2EVarNet)~\cite{sriram2020end} by incorporating a second
learned term in the unrolled iterates that injects guide information.
Starting from the zero-filled reconstruction
${\x^{(0)} = \ObsOp^\herm\y}$, each layer implements an unrolled projected gradient-descent update,
\begin{equation}\label{eq-tgvn}
\begin{aligned}
    \x^{(k+1)} =\,& \x^{(k)}
    - \eta^{(k)} \ObsOp^\herm(\ObsOp\x^{(k)} - \y) \\
    &- \mu^{(k)} \mathcal{P}_\delta\, \br^{(k)}
    - g(\x^{(k)}; \bphi^{(k)}),
\end{aligned}
\end{equation}
where $g$ is a learned regularizer carrying population priors with parameters
$\bphi^{(k)}$, and $\mathcal{P}_\delta$ is the orthogonal projector onto the
span of the right singular vectors of $\ObsOp$ whose singular values fall below
a learned threshold $\delta$ (the ``ambiguous space,'' e.g., unacquired k-space in the single-coil case
and high $g$-factor directions in the multi-coil case). The two methods differ
in how the guide residual $\br^{(k)}$ is formed. TGVN maps the guide alone,
${\br^{(k)} = \x^{(k)} - f(\w; \btheta^{(k)})}$, whereas L-TGVN conditions the
guidance network on the current iterate as well,
${\br^{(k)} = \x^{(k)} - f(\w,\, \x^{(k)};\, \btheta^{(k)})}$, with $f$ a
dual-encoder, shared-decoder U-Net~\cite{unet} reading the prior-scan slices and
the current estimate on separate encoders.
Like E2EVarNet, L-TGVN estimates the coil sensitivities with another U-Net.

Although E2EVarNet and (L-)TGVN incorporate the MRI encoding operator through
explicit data-consistency updates, their learned regularization relies on
``black-box'' neural network architectures (such as U-Nets), largely opaque
to traditional analysis, rather than regularizers explicitly derived from
signal-processing models~\cite{monga2021unrolling, janjusevicCDLNet2022,
janjusevic2025groupcdl}.
In L-TGVN, a learned guidance network uses both the prior images and the current
reconstruction to implicitly account for inter-visit misregistration,
acquisition differences, and image phase variation. The trust projection, a key
component of L-TGVN, further restricts the influence of this guidance primarily to
the ambiguous space of the MRI encoding operator, therefore reducing the risk of
violating data consistency. However, since L-TGVN relies on the learned guidance
network to determine how prior information should be used, it does not provide
an explicit measure of local feature correspondence or a clear indication of
which prior features contribute to the reconstruction.

In contrast, LGGS explicitly incorporates prior-image information into the
reconstruction regularizer through guided group sparsity, rather than relying on
a separate learned guidance network. A novel phase-invariant circulant attention
mechanism is used to identify corresponding features between the current and
prior exams, despite residual spatial misregistration and inter-visit phase
differences (Section~\ref{sec-circatt}). The similarity between current and
prior features determines the extent to which prior information should influence
reconstruction regularization. Importantly, the LGGS architecture explicitly
prevents the direct transfer of prior-image features into the reconstruction.
Instead, prior information influences the reconstruction through learned group
thresholding, controlling the suppression or preservation of current
reconstruction features, while standard data-consistency updates are enforced
based on the acquired k-space data. This provides an explicit and interpretable
mechanism for incorporating longitudinal information into reconstruction
regularization, in contrast to existing approaches that rely on implicitly
learned guidance.

\begin{figure*}[tbh]
    \centering
    \includegraphics[width=0.8\linewidth]{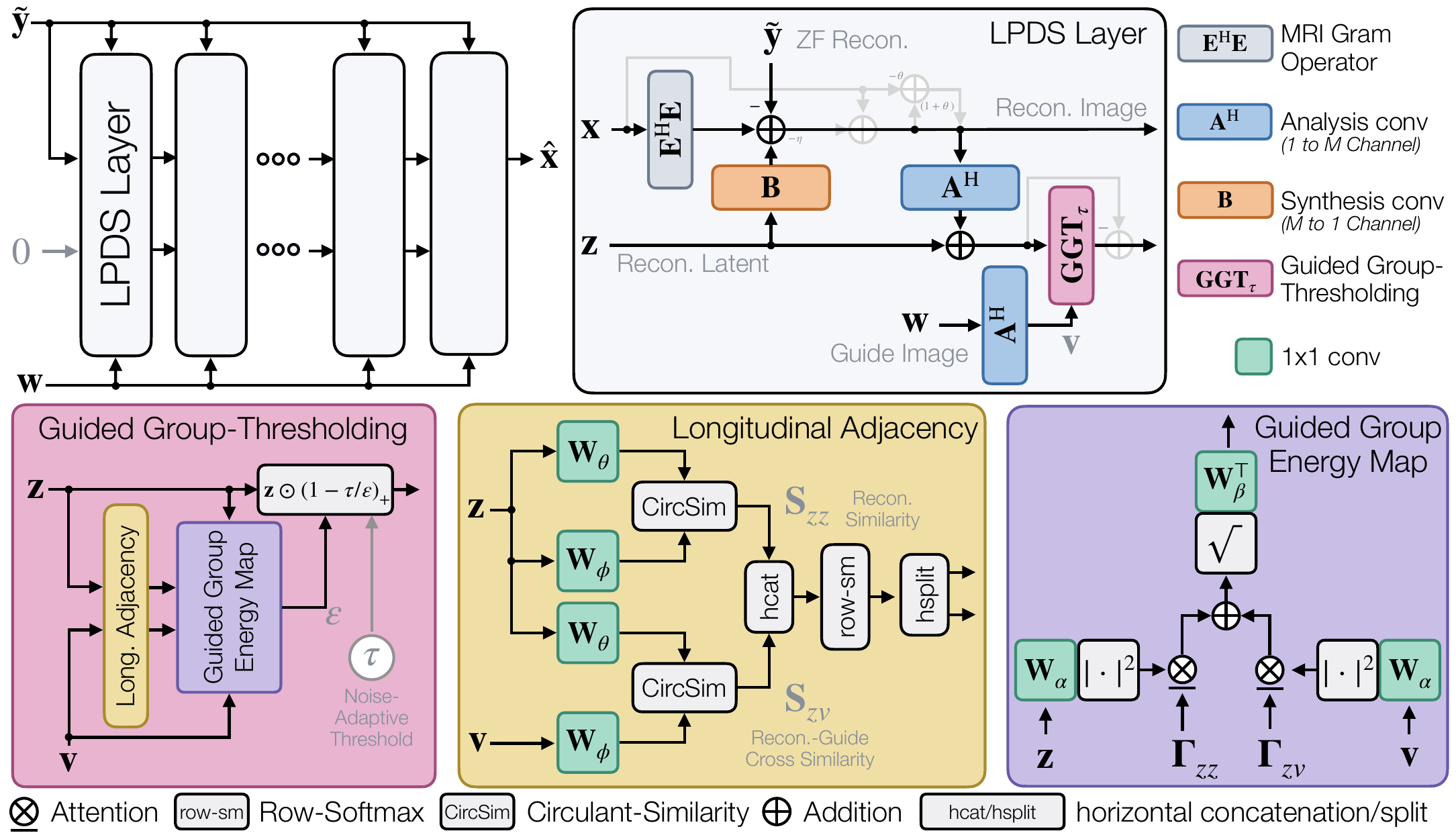}
    \caption{%
        LGGS architecture. $K$ unrolled primal--dual iterations \eqref{eq-lggs}.
        The guide image $\w$ enters only through its analysis codes $\bv$, which
        reach the dual update through the guided group-thresholding operator
        $\GGT_{\btau}(\z;\,\bv)$ \eqref{eq-ggt}, whose threshold is divided by
        the guided group energy map
        $\bEnergy^{\alpha\beta}_{\ADJMAT}(\z;\,\bv)$
        \eqref{eq-guided-energy-ab}, in which the longitudinal adjacency
        $\ADJMAT_{zv}$ acts. The subtraction following $\GGT_{\btau}$ forms its
        Fenchel conjugate $\GGT^\ast_{\btau}$. Setting $\bv = 0$ recovers the
        unguided LPDSNet~\cite{lpds2025} exactly.
    }
    \label{fig-arch}
\end{figure*}
\subsection{Learned Group Sparsity}\label{sec:groupcdl}

GroupCDL~\cite{janjusevic2025groupcdl} introduced a unified deep-learning
framework that incorporates local and non-local image priors into MRI
joint denoising and reconstruction (JDR). The GroupCDL architecture derives
from the generalized group-sparsity penalized synthesis-dictionary
sparse-coding objective,
\begin{equation} \label{eq-synthesis-sc}
\z^\diamond = \argmin_{\z} \frac{1}{2}\norm{\y - \ObsOp\bD\z}_2^2
            + \lambda \psi(\z),
\end{equation}
where ${\hat{\x} = \bD\z^\diamond}$ and the group-sparsity penalty is,
\begin{equation} \label{eq-group-sparsity}
    \psi(\z) = \sum_{i=1,m=1}^{Q,M} \sqrt{\sum_{j=1}^{Q} \ADJMAT_{ij}\, \abs{\z_m[j]}^2}
             = \norm{\sqrt{\cwise{\ADJMAT}\, \abs{\z}^2}}_1.
\end{equation}
Here, ${\ADJMAT \in \R_+^{Q\times Q}}$ is a (soft) adjacency matrix encoding
group membership over the $Q$ pixels of the $M$-feature latent representation
${\z \in \C^{QM}}$ of the reconstruction: entry $\ADJMAT_{ij}$ reflects the
degree to which pixel $j$ belongs to the group of the query pixel $i$, and the
same adjacency is shared across all $M$ feature channels. This adjacency
matrix is formed dynamically within each GroupCDL unrolling step. For
computational tractability, the number of nonzero entries per row (the
``window size'') is limited; when reduced to 1, the penalty simplifies to
$\ell_1$ sparsity, ${\psi(\z) = \norm{\z}_1}$.

GroupCDL unrolled \eqref{eq-synthesis-sc} via proximal-gradient descent, where
${\prox_\psi(\z) = \argmin_{\bv}\,\psi(\bv) + \tfrac{1}{2}\norm{\bv-\z}_2^2}$
is approximated by the \emph{Learned Group-Thresholding} (LGT) module,
\begin{equation}\label{eq-lgt}
    \GT_{\btau}(\z) = \z \odot
        \Bigg(1 - \frac{\btau}{\pwise{\bW}_\beta^\top\!
            \sqrt{\cwise{\ADJMAT}\abs{\pwise{\bW}_\alpha\z}^2}
        }\Bigg)_{\!\!+},
\end{equation}
with learned parameters $\bW_\alpha,\,\bW_\beta,\,\btau$ and where
${(\cdot)_+}$ denotes projection onto the non-negative orthant $\R_+$.

The adjacency matrix ${\ADJMAT \in \R_+^{Q \times Q}}$ is formed dynamically at each layer from the
current latent codes using a circulant-attention similarity
metric~\cite{janjusevic2025groupcdl}.
Specifically,
\begin{align}
    \ADJMAT &= \rowsoftmax(\circsim(\pwise{\bW_\theta}\z,\,\pwise{\bW_\phi}\z)), 
\end{align}
where ${\bW_\theta,\bW_\phi \in \R^{M_h \times M}}$ are learned matrices, and 
$\SIMMAT = \circsim(\bu, \bv) \in \R^{Q\times Q}$ is the $QW^2$-sparse
sliding-window circulant similarity matrix, 
\begin{align} \label{eq-similarity-gcdl}
    \SIMMAT_{i,j} = 
    \begin{cases}
        -\frac{1}{2}\norm{\bu[i] - \bv[j]}_2^2, & j \in \text{window}(i, W) \\
        -\infty, &\text{otherwise}
    \end{cases},
\end{align}
where ${W \geq 1}$ is the similarity search window size and $\rowsoftmax$ is the
row-wise soft-max function which normalizes the similarities within each search
window (${\rowsoftmax(\SIMMAT)_{ij} = e^{\SIMMAT_{ij}}/(\sum_{j} e^{\SIMMAT_{ij}})}$). The $i$-th row of $\ADJMAT$ thus encodes the relative similarity of
latent vectors $\z[j]$ to $\z[i]$ for every latent pixel. By forming $\ADJMAT$
dynamically within the network, GroupCDL enables group sparsity to be learned
end-to-end, whereas classical group sparsity typically requires
hand-crafted precomputation of $\ADJMAT$~\cite{mairal2009non}.
The learned matrices ${\bW_{\theta}, \bW_\phi, \bW_\alpha, \bW_\beta \in
\R^{M_h \times M}}$ are introduced simply to reduce the computational
complexity of the attention mechanism by setting ${M_h \ll M}$, which we refer
to as ``feature-compression''.

GroupCDL was demonstrated to achieve \soa supervised MRI JDR performance with a
fraction of the learned parameters of
E2EVarNet~\cite{janjusevic2025groupcdl}. GroupCDL employs group sparsity to
encode a signal-processing derived non-local self-similarity prior over a single
image. In this work we employ it to form a prior which operates non-locally in
time (longitudinally), regularizing the reconstruction under the assumption that similar image
regions at different time points share feature-domain sparsity patterns.

\subsection{Learned Primal-Dual Splitting}\label{sec:lpdsnet}

LPDSNet~\cite{lpds2025} introduced an interpretable MRI reconstruction network
derived from unrolling the \emph{analysis-dictionary} convolutional sparse-coding
objective,
\begin{equation} \label{eq-analysis-sc}
\min_{\x} \frac{1}{2}\norm{\y - \ObsOp\x}_2^2 + \psi(\bD^\herm\x),
\end{equation}
with $\psi = \norm{\cdot}_1$.
This objective was unrolled for $K$ iterates via primal-dual splitting
(PDS)~\cite{chambolle_ctsopt} and direct parameterization of $\bD^\herm$: each
layer learns an analysis convolution ${\bA^\herm}$ ${(1 \rightarrow M)}$, a
synthesis convolution $\bB$ ${(M \rightarrow 1)}$, step-sizes
${\eta, \theta \in \R_+}$, and the dual proximal operator, which for
${\psi = \norm{\cdot}_1}$ is an element-wise clipping with learned
feature-dependent thresholds. The guided unrolling \eqref{eq-lggs} of
Section~\ref{sec:methods} is this same iteration, differing only in that
proximal operator.

LPDSNet demonstrated superior performance to the local (non-group)
counterpart of GroupCDL, CDLNet~\cite{janjusevicCDLNet2022}, on both supervised
and unsupervised MRI JDR tasks~\cite{lpds2025}. The analysis-sparsity
formulation is particularly attractive for the longitudinal setting considered
here, as the guide image's analysis-domain codes can be obtained by a single analysis
convolution $\bv = {{\bA}^{(k)}}^\herm \w$, in contrast to the synthesis model of
GroupCDL, which may require a separate sparse-coding pass to obtain an
equivalent synthesis-domain latent representation.

\section{Proposed Method}\label{sec:methods}
Our proposed framework, LGGS, derives longitudinal guidance from the
group-sparsity prior \eqref{eq-group-sparsity}, generalized to admit a guide
and unrolled as a primal-dual splitting solver. Figure~\ref{fig-arch} gives the
full architecture in block-diagram form.

\subsection{Longitudinal Guided Group Sparsity (LGGS)}
\label{sec:methods-reg}
Let ${\bv \in \C^{QM}}$ denote the analysis codes of the guide
image $\w \in \C^N$. We define the \emph{guided group-sparsity} regularizer as,
\begin{align}\label{eq-guided-psi}
    \psi(\z;\,\bv)
        &= \norm{\sqrt{
            \cwise{\ADJMAT}_{zz}\abs{\z}^2 + \cwise{\ADJMAT}_{zv}\abs{\bv}^2
        }}_1,
\end{align}
where ${\ADJMAT = \begin{bmatrix} \ADJMAT_{zz} & \ADJMAT_{zv}\end{bmatrix} \in
\R_+^{Q\times 2Q}}$ is row-normalized, i.e.\ $\norm{\ADJMAT_{i:}}_1 = 1$ for all
$i$. The block ${\ADJMAT_{zv}\in\R_+^{Q\times Q}}$ carries the cross-similarity
between the reconstruction and guide codes. The block $\ADJMAT_{zz}$ is \emph{diagonal by construction}, LGGS searching a
single pixel for self-similarity against a $W_g \times W_g$ neighborhood for
guide similarity (Section~\ref{sec-circatt}). This is the defining property of
the proposed regularizer: \emph{every off-diagonal entry of $\ADJMAT$
constitutes a reconstruction-to-guide coupling}. LGGS therefore introduces no
intra-image non-local prior, and all of its grouping acts along the longitudinal
dimension. 

The regularizer is minimized over $\z$ alone. 
Absent a guide, the regularizer degenerates to
unstructured sparsity $({\psi(\z;\,\bm{0}) = \norm{\z}_1})$, and the resulting
unrolling degenerates to LPDSNet~\cite{lpds2025}. 

\subsection{Guided Reconstruction with Learned Primal-Dual Splitting}
\label{sec-unrolling}
We unroll primal-dual splitting for $K$ iterations with the guided prior~\eqref{eq-guided-psi}.
First, we define the \emph{guided $\ADJMAT$-energy map} and its feature-compressed form as,
\begin{align}
    \bEnergy_{\ADJMAT}(\z;\bv)
        &\triangleq \sqrt{\cwise{\ADJMAT}_{zz}\,\abs{\z}^2
                         + \cwise{\ADJMAT}_{zv}\,\abs{\bv}^2}, \label{eq-guided-energy}\\
    \bEnergy^{\alpha\beta}_{\ADJMAT}(\z;\bv)
        &= \bigl(\pwise{\bW_\beta}^\top \circ \bEnergy_{\ADJMAT}
           \circ \pwise{\bW_\alpha}\bigr)(\z; \bv), \label{eq-guided-energy-ab}
\end{align}
with $\circ$ function composition and $\pwise{\bW}$ applied to $\z$ and $\bv$
alike. The \emph{Guided Group-Thresholding} (GGT) proximal operator is then,
\begin{equation}\label{eq-ggt}
    \GGT_{\btau}(\z;\,\bv) = \z \odot \left(1 -
        \frac{\btau}{\bEnergy^{\alpha\beta}_{\ADJMAT}(\z;\,\bv)}\right)_{\!\!+},
\end{equation}
with Fenchel conjugate
${\GGT^\ast_{\btau}(\z;\,\bv) = \z - \GGT_{\btau}(\z;\,\bv)}$; setting
${\bv = 0}$ recovers the LGT module \eqref{eq-lgt}. The full LGGS unrolling
proceeds as,
\begin{equation}\label{eq-lggs}
\begin{gathered}
    \tilde{\y} = \ObsOp^\herm\y - \ObsOp^\herm\ObsOp(\mu\bOne),\quad
    \x^{(0)} = \tilde{\y},\quad
    \z^{(0)} = 0, \\ %
    \text{\small for $k=0:K-1$} \hspace{0.6\columnwidth} \\
    \begin{array}{|rcl}
        \x^{(k+1)} &=& \x^{(k)} - \eta\!\left(
            \ObsOp^\herm\ObsOp\x^{(k)} - \tilde{\y} + \bB\z^{(k)}\right) \\[4pt]
        \tilde{\x}  &=& \x^{(k+1)} + \theta\!\left(\x^{(k+1)} - \x^{(k)}\right) \\[4pt]
        \z^{(k+1)} &=& \GGT^\ast_{\btau}\!\left(\z^{(k)} + \bA^\herm\tilde{\x} ;\, \bA^\herm (\w - \mu_{\w}) \right)
    \end{array}\\
    \hat{\x} = f_\Theta(\y;\,\ObsOp,\,\w) = \x^{(K)} + \mu,
\end{gathered}
\end{equation}
where per-layer parameters $\eta,\,\theta,\,\bA,\,\bB,\,\btau$ are learned
(layer index suppressed for readability), $\mu$ and $\mu_{\w}$ are the means of
the zero-filled reconstruction and of the guide, and $\bOne$ is the image of all
ones (see Appendix~\ref{app:mean-subtraction} for a derivation of operator-aware mean-subtraction).

\subsection{Circulant Attention for the Adjacency Matrix}
\label{sec-circatt}

As in GroupCDL (Section~\ref{sec:groupcdl}), the adjacency matrix $\ADJMAT$ is
formed dynamically within the network. LGGS now
extends the adjacency column-wise to carry the reconstruction--guide cross-similarity,
\begin{align}
    \ADJMAT &= \rowsoftmax\!\left(
        \begin{bmatrix}\SIMMAT_{zz} & \SIMMAT_{zv}\end{bmatrix}
    \right), \label{eq-adjmat}\\
    \SIMMAT_{zz} &= \circsim(\pwise{\bW_\theta}\z,\,\pwise{\bW_\phi}\z),
                   \label{eq-simzz}\\
    \SIMMAT_{zv} &= \circsim(\pwise{\bW_\theta}\z,\,\pwise{\bW_\phi}\bv),
                   \label{eq-simzv}
\end{align}
with $\circsim$ the circulant similarity matrix of \eqref{eq-similarity-gcdl}.

The two blocks are assigned \emph{separate} search windows, by which the
longitudinal restriction of Section~\ref{sec:methods-reg} is imposed. The
self-similarity block $\SIMMAT_{zz}$ is computed with window ${W_z = 1}$ and the
guide block $\SIMMAT_{zv}$ with ${W_g > 1}$, so that $\ADJMAT$ stores longitudinal adjacency only. 
In this manuscript, we only consider longitudinal group sparsity and leave the
investigation of a joint spatial-longitudinal prior to future work. 

This construction is translation-equivariant by design, making
$\ADJMAT$ inherently robust to local misalignment between the
reconstruction and the guide. Global displacement is removed beforehand by the
integer-voxel translation of Section~\ref{sec:setup}, leaving the adjacency responsible
for the spatially varying local misalignment which a rigid shift cannot
represent.
Unlike in GroupCDL, $\ADJMAT$ is never formed explicitly. The
energy map \eqref{eq-multi-energy} is instead evaluated by the novel fused
circulant flash-attention kernels of
CirculantAttention.jl\footnote{\href{https://github.com/nikopj/CirculantAttention.jl}{https://github.com/nikopj/CirculantAttention.jl}}. 

\subsection{Phase-Invariant Similarity}\label{sec-phase-invariant}
The similarity function $\simfun$ of \eqref{eq-similarity-gcdl} compares local
feature vectors of the reconstruction and guide. In the longitudinal setting the
standard real inner-product over ${\C^M \cong \R^{2M}}$ is poorly suited to this:
prior and target examinations of the same anatomy differ by a spatially varying
phase, arising from magnetic field inhomogeneity, coil loading, and coil
sensitivity estimation~\cite{robinson2017phase}, which carries little structural meaning. Two
patches of identical magnitude content separated by a relative phase $\varphi_0$ yield
a real inner-product scaling as $\cos(\varphi_0)$, which vanishes near $\varphi_0 = \pi/2$
and changes sign beyond it.

We therefore define a \emph{phase-invariant} similarity by maximizing over a
single scalar rotation per query--key pair,
\begin{equation}\label{eq-phase-invariant-def}
    \pisim(\bu,\bv) \,\triangleq\, \max_{\varphi \in [0,2\pi)}~
    \simfun\!\left(\bu,\, e^{j\varphi}\bv\right),
    \qquad \bu,\bv \in \C^M,
\end{equation}
which is solved in closed form. The dot-product similarity gives
${\pidotsim(\bu,\bv) = \abs{\bv^\herm\bu}}$, the magnitude of the complex inner
product, and the negative distance similarity gives the \emph{phase-invariant
distance similarity},
\begin{equation}\label{eq-pidistsim}
    \pidistsim(\bu,\bv) =
        -\tfrac{1}{2}\!\left(\norm{\bu}_2^2
        - 2\abs{\bv^\herm\bu} + \norm{\bv}_2^2\right).
\end{equation}
Neither collapses to a magnitude-only metric of the form
$\abs{\bv}^\top\!\abs{\bu}$, which would discard the structure carried by sign
and phase within each feature vector and so match an edge patch to a constant
one of equal magnitude. We use $\pidistsim$ for every cross-similarity term
involving the guide, so that $\ADJMAT_{zv}$ couples reconstruction and guide
features on phase-invariant structural agreement alone. Derivations and the
back-propagation rules required to train through $\pidistsim$ are given in
Appendix~\ref{app:phase-invariant}, and Section~\ref{sec:res-abl-guidesim} validates the choice
experimentally.

\subsection{Multiple Guide Images} \label{sec-mglggs}
As the circulant-attention mechanism requires no slice-to-slice correspondence
beyond the global alignment of Section~\ref{sec:setup}, the LGGS framework
extends naturally to multiple guide images ($N_g \geq 1$).
We extend the adjacency matrix column-wise to include cross-similarity terms
for each guide, yielding the \emph{multi-guide energy map},
\begin{align}
    \bEnergy_{\ADJMAT}\!\left(\z;\,\{\longi{\bv}{g}\}_{g=1}^{N_g}\right)
        &\triangleq \sqrt{
            \cwise{\ADJMAT}_{zz}\,\abs{\z}^2
            + \sum_{g=1}^{N_g} \longij{\cwise{\ADJMAT}}{(g)}{zv}\,
              \abs{\longi{\bv}{g}}^2
        }. \label{eq-multi-energy}
\end{align}
Note that \eqref{eq-multi-energy} reduces to the single-guide energy
\eqref{eq-guided-energy} when $N_g = 1$, and further to the unguided energy
$\norm{\z}_1$ when $N_g = 0$ (empty sum). The learned multi-guide group-thresholding operator then takes the same form as
the single-guide case \eqref{eq-ggt}, with $\bEnergy^{\alpha\beta}_{\ADJMAT}$
evaluated on the guide set.
The extended adjacency matrix is formed from the concatenated similarity
matrices,
\begin{equation}\label{eq-multi-guide-adjmat}
\begin{aligned}
    \ADJMAT &=
        \begin{bmatrix}
            \ADJMAT_{zz}
            & \longij{\ADJMAT}{(1)}{zv}
            & \cdots
            & \longij{\ADJMAT}{(N_g)}{zv}
        \end{bmatrix} \\
    &= \rowsoftmax\!\left(
        \begin{bmatrix}
            \SIMMAT_{zz}
            & \longij{\SIMMAT}{(1)}{zv}
            & \cdots
            & \longij{\SIMMAT}{(N_g)}{zv}
        \end{bmatrix}\right),
\end{aligned}
\end{equation}
with $\SIMMAT_{zz}$ as in \eqref{eq-simzz} and each
$\longij{\SIMMAT}{(g)}{zv}$ formed as in \eqref{eq-simzv} from the $g$-th
guide's codes $\longi{\bv}{g}$.
Row-normalizing $\ADJMAT$ in \eqref{eq-multi-guide-adjmat} keeps the total group
energy normalized regardless of $N_g$, so that $\btau$ needs no recalibration as
the number of guides changes. We take the $N_g$ slices of the prior-scan volume
centered on the target slice. A target lying between two guide slices then draws
on both, with weights set by their reconstruction-space similarity, which is what
confers robustness to out-of-plane misregistration.
In Section~\ref{sec:res-abl-nguides}, we experimentally validate the benefit of
using multiple slices from the prior scan to guide LGGS-based
reconstruction.

\section{Experimental Results}\label{sec:results}
\subsection{Experimental Setup}\label{sec:setup}
\noindent\textbf{Dataset}:
We used an in-house dataset of serial brain MRI exams acquired from $70$
clinical patients with brain tumors, each of whom underwent three MRI exams.
The data were collected under institutional review board (IRB) protocol
\mbox{i25-00504} and used retrospectively for this study, with a waiver of informed
consent.
Fully sampled, T1-weighted multi-coil raw k-space data were available for the
exams, providing complex-valued MRI information that preserves both magnitude
and phase. The dataset comprised $210$ volumes with in-plane dimensions of
$256\times 256$ and approximately $50$--$80$ slices per volume. Coil arrays and
acquisition protocols varied across patients and, in some cases, between visits
of the same patient. Twelve patients ($36$ volumes) were held out for
independent testing, and the remaining $58$ patients were used for training,
with $5\%$ of the training volumes reserved for validation. No patient appeared
in more than one split. For evaluation, each patient contributed three target
volumes for reconstruction. Visit~1 was reconstructed using Visit~2 as the
guide, Visit~2 using Visit~1, and Visit~3 using Visit~2. For training, all
ordered pairs of available exams from each patient were used. Reconstruction
was performed slice by slice in the axial plane.

\noindent\textbf{Data Preparation}:
For each subject, the three volumes were aligned and registered to Visit~1
using an integer-voxel translation estimated from low-resolution coil-combined
images reconstructed from the autocalibration signal (ACS) data. This
corrected global in-plane and through-plane misregistration without
interpolation. Because the registration requires only the ACS data, it can also
be performed at inference time. A per-slice brain mask derived from ESPIRiT
sensitivity maps of the fully sampled volume was used to suppress background in
the reconstruction target and determine the display range for qualitative
figures. Each visit was then normalized using a robust scalar estimated from
the ACS data. The complete data preparation pipeline is described in
Appendix~\ref{app:data-prep}.

\noindent\textbf{Undersampling}:
Retrospective undersampling was performed using a one-dimensional Cartesian
sampling pattern with equispaced phase-encoding lines. The fully sampled ACS
region was added to the sampling pattern
(Table~\ref{tab:accel}). Because the ACS lines were added rather than
substituted for the equispaced lines, the effective acceleration rate
($R_\mathrm{eff}$) was lower than the nominal rate ($R$), against which the rate
sweeps of Section~\ref{sec:res-main} are plotted. For LPDSNet and LGGS, coil
sensitivity maps were estimated using ESPIRiT~\cite{espirit} from the
Hamming-windowed ACS data, with a $2\times 2$ calibration kernel and no
eigenvalue thresholding. The minimum ACS region was set to $2.67\%$ to
maintain an overdetermined calibration system and improve the stability of the
eigenvalue decomposition. 

\begin{table}[tbh]
\centering
\caption{Uniform undersampling configuration. The fully sampled center shrinks
    with the nominal rate $R$ until ${R = 12}$.}
\label{tab:accel}
\vspace{0.3em}
\begin{tabular}{l|ccccc} \hline
    nominal rate $R$      & $6\times$ & $8\times$ & $10\times$ & $12\times$ & $16\times$ \\ \hline
    center-fraction (\%)    & 5.33 & 4.00 & 3.20 & 2.67 & 2.67 \\
    $R_\mathrm{eff}$      & 4.74  & 6.40  & 8.00  & 8.83  & 11.64 \\ \hline
\end{tabular}
\end{table}

\begin{table}[tbh]
\centering
\caption{Learned parameter count, and wall-clock time of one
forward and backward pass over a complete reconstruction on a single NVIDIA
A100. ESPIRiT is coil-map estimation rather than a reconstruction method: it
carries no learned parameters, is never backpropagated, and runs once per slice
to build the forward operator given to LPDSNet and LGGS.}
\label{tab:bench}
\vspace{0.3em}
\begin{tabular}{l|cccc} \hline
    \multirow{2}{*}{Method} & \multirow{2}{*}{params} & \multicolumn{3}{c}{time (ms)} \\ \cline{3-5}
              &          & forward & backward & total \\ \hline
    E2EVarNet & $33.3$\,M & $81$  & $410$ & $491$ \\
    LPDSNet   & $1.00$\,M & $85$  & $155$ & $241$ \\
    L-TGVN    & $43.6$\,M & $216$ & $620$ & $836$ \\
    LGGS      & $1.08$\,M & $234$ & $547$ & $780$ \\
    ESPIRiT   & --        & $166$ & --    & $166$ \\ \hline
\end{tabular}
\end{table}
\noindent\textbf{Architecture}:
LGGS and LPDSNet shared every hyperparameter except guidance.
The LGGS self-similarity window was ${W_z=1}$, so $\ADJMAT_{zz}$ was diagonal and
\emph{all} non-local coupling in LGGS was longitudinal
(Section~\ref{sec-circatt}); the guide window was ${W_g=3}$ over $N_g=3$
neighboring slices of the guide visit, compared under $\pidistsim$
\eqref{eq-pidistsim}, unless otherwise mentioned. Filters were complex-valued,
and $\bW_{\{\theta,\phi,\alpha,\beta\}}$ were tied across layers. Both LGGS and
LPDSNet used coil sensitivity maps estimated from the ACS data with ESPIRiT.
E2EVarNet and L-TGVN estimated coil sensitivity maps directly from the ACS data
using learned sensitivity networks, following their original implementations.
E2EVarNet used $12$ cascades, while L-TGVN used $6$ cascades and the same three
prior guide slices (${N_g=3}$) as LGGS, supplied as magnitude images. The network
capacity was comparable between E2EVarNet and L-TGVN, as shown in
Table~\ref{tab:bench}.

The L-TGVN implementation was adapted from the source code released by the
original authors. The original L-TGVN study had ${\sim}500$ million parameters
and used eight NVIDIA A100 GPUs for training. Due to computational resource
constraints, we used a reduced-capacity configuration with $43.6$ million
parameters. We acknowledge that a larger network capacity may further improve
L-TGVN performance, but evaluating the original full-scale configuration was
beyond the computational resources available for this study. Complete
architecture details are provided in Appendix~\ref{app:arch}.

\noindent\textbf{Training}:
All four models were trained identically: 
$3\times 10^{5}$ gradient steps at batch size $1$ on whole
slices, Adam~\cite{adam} with cosine annealing from $4\times 10^{-4}$ to
$2\times 10^{-6}$ and gradient-norm clipping at $1$.
A supervised normalized $\ell_1$-$\ell_2$ loss was used on the brain-masked
reconstruction magnitude.

\noindent\textbf{Metrics}:
NRMSE, SSIM, and LPIPS~\cite{zhang2018perceptual} (AlexNet backbone) were
computed between magnitude images, on pixels within the brain-support mask only. 
Appendix~\ref{app:unmasked} shows that metrics without the 
brain-support mask become artificially inflated. 

\noindent\textbf{Implementation}:
All models were implemented in the Julia~\cite{julia} programming language with
Lux.jl~\cite{pal2023lux}. 

\noindent\textbf{Hardware}:
All models were trained on a single NVIDIA A100 GPU with eight CPU cores.

\noindent\textbf{Statistical Reporting}:
Unless noted otherwise, every quantitative figure below reports means over 36
volumes from 12 held-out patients, and error bars and shaded bands give the
standard error taken with the 12 patients as the independent unit. Significance
was assessed by a two-sided exact Wilcoxon signed-rank test on the 12 paired
patient means.
The bar charts mark the test above
each bar, $\ast\ast$ or $\square\square$ for $p \le 0.001$ and $\ast$ or $\square$ for $p \le 0.01$, with no marker
above $0.01$. The rate plots carry the same test in the marker itself: a \emph{filled} marker is
$p \le 0.001$, an \emph{open} marker is $p \le 0.05$, and a \emph{gray} marker is
not significant at $0.05$. 

\begin{figure}[tbh]
    \centering
    \includegraphics[width=\columnwidth]{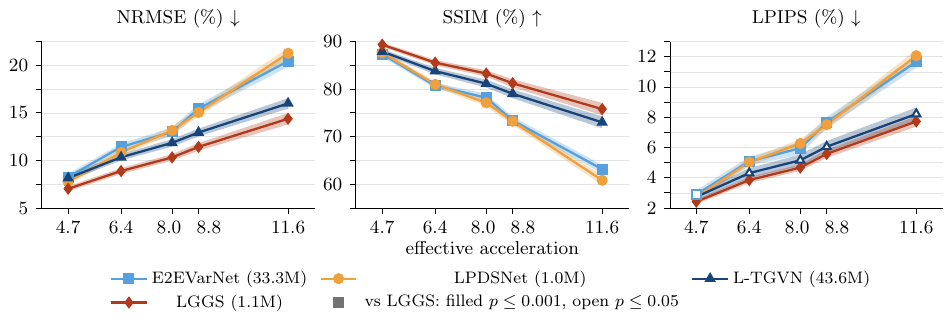}
    \caption{%
        Reconstruction quality across effective acceleration rates (see Table~\ref{tab:accel}) on the longitudinal
        brain dataset, for the unguided LPDSNet~\cite{lpds2025} and
        E2EVarNet~\cite{sriram2020end}, the guided longitudinal
        L-TGVN~\cite{ltgvn2026}, and the proposed LGGS, over three metrics
        (NRMSE\,/\,SSIM\,/\,LPIPS).
    }
    \label{fig-results-accel}
\end{figure}
\begin{figure*}[tbh]
    \centering
    \includegraphics[width=\linewidth]{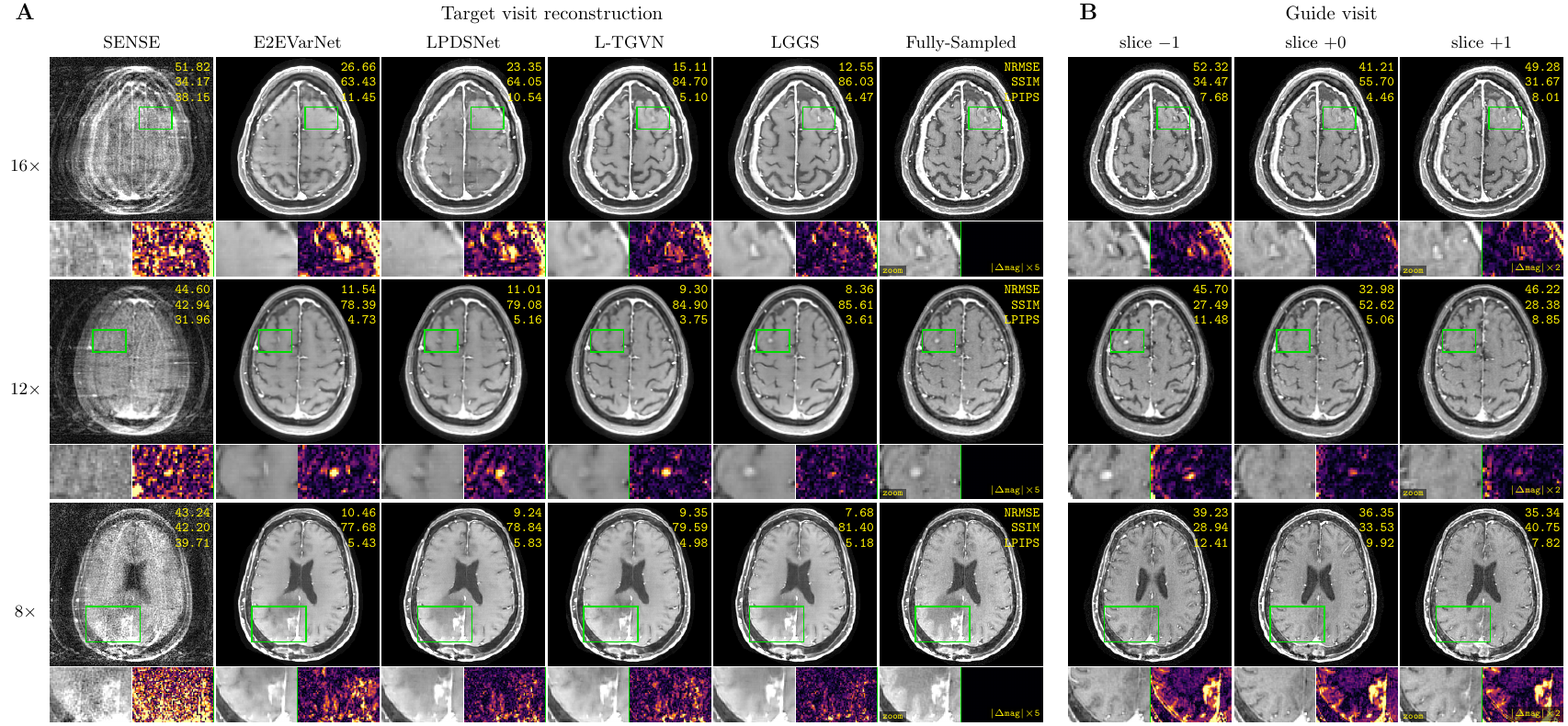}
    \caption{%
        Visual comparison, one case per row, each shown at its own acceleration
        rate. \textbf{(A)} Magnitude reconstructions of the target visit, with
        NRMSE\,/\,SSIM\,/\,LPIPS ($\times 100$) above each panel.
        \textbf{(B)} Magnitude images of the prior visit supplied to the guided methods (3 slices), with
        \emph{residuals and metrics taken against the target visit} at $\times 2$ gain.
        Slice $+0$ denotes the center guide slice for the corresponding target slice.
        Every panel within a row crops the same pixel window.
    }
    \label{fig-results-compare}
\end{figure*}
\begin{figure}[tbh]
    \centering
    \includegraphics[width=0.94\columnwidth]{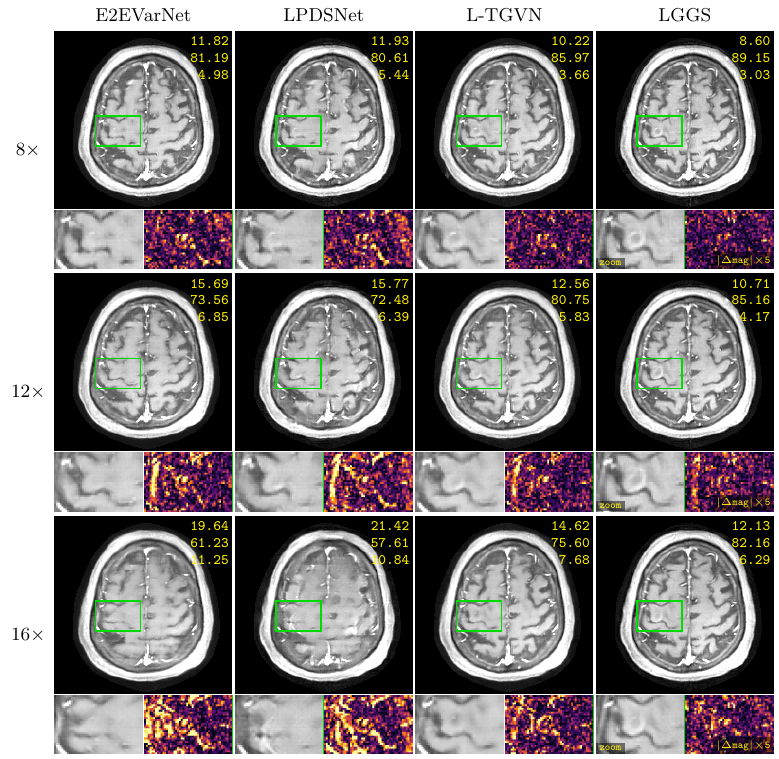}
    \caption{%
        A single slice reconstructed across acceleration rates, with
        every panel cropped to the same window. NRMSE\,/\,SSIM\,/\,LPIPS
        ($\times 100$) are given above each panel.
    }
    \label{fig-results-accelrows}
\end{figure}
\subsection{Performance Across Acceleration Rates}
\label{sec:res-main}
In this section we compared LGGS against its unguided counterpart
LPDSNet~\cite{lpds2025}, the unguided E2EVarNet~\cite{sriram2020end},
and the longitudinally guided L-TGVN~\cite{ltgvn2026}, over the five rates
of Table~\ref{tab:accel}. 
Figure~\ref{fig-results-accel} shows quantitative reconstruction results with
each method from $6\times$ to $16\times$ nominal uniform acceleration rate. We
observed that LGGS attained the best NRMSE, SSIM, and LPIPS at every rate, with every difference significant. 
The two unguided networks performed comparably to one another and degraded the
fastest, with the guided to unguided gap widening considerably at the higher
nominal acceleration rate of $16\times$. As LPDSNet differs from LGGS only in
the guide block of \eqref{eq-guided-psi}, this separation is attributable to
the longitudinal prior alone. We observed that it widened monotonically with
acceleration rate, across all metrics.

Against L-TGVN, which carries the same prior examination through a learned
guidance network, LGGS improved NRMSE by between $1.12$\% and $1.62$\% across the
sweep, at $p \le 0.001$ on NRMSE and SSIM at every rate; only LPIPS softened, to
$p \le 0.01$ at intermediate rates and $p \le 0.05$ at $11.64\times$. The same
curves may also be read at matched error rather than at matched rate: LGGS
attained $10.32$\% NRMSE at $10\times$ nominal, whereas L-TGVN required $8\times$ nominal to
attain $10.33$\% NRMSE, affording LGGS a $25\%$ increase in acceleration at equal error.

The comparison was matched on computation rather than on parameters
(Table~\ref{tab:bench}). Within each family the parameter counts were comparable,
$1.00$M against $1.08$M for LPDSNet and LGGS and $33.3$M against $43.6$M for
E2EVarNet and L-TGVN, while across families they differed more than thirty-fold.
Wall-clock cost was far closer: one forward and backward pass took LGGS $780$\,ms
against L-TGVN's $836$\,ms. Map estimation added a further $166$\,ms to
the LPDSNet and LGGS pipelines, which the variational-net baselines absorbed into their
own sensitivity networks, so the guided comparison was made at approximately
equal cost despite the $40\times$ difference in parameter count.

Figures~\ref{fig-results-compare} and~\ref{fig-results-accelrows} highlight the
qualitative differences behind these metrics. We observed that the residual maps
of the unguided reconstructions carried visible anatomical structure which grew
with acceleration rate, whereas those of LGGS remained comparatively
unstructured. At $16\times$ nominal acceleration (Figure~\ref{fig-results-accelrows}, bottom row),
E2EVarNet and LPDSNet retained markedly less detail in the magnified region than
LGGS. Part~B of Figure~\ref{fig-results-compare} provides the prior examination
supplied to each guided method, with residuals and metrics taken against the \emph{target},
so that the inter-visit change carried by the guide may be both quantitatively and qualitatively assessed alongside the
reconstruction error. We observed that LGGS resolved
important pathological features more consistently than L-TGVN.

\begin{figure}[tbh]
    \centering
    \includegraphics[width=\columnwidth]{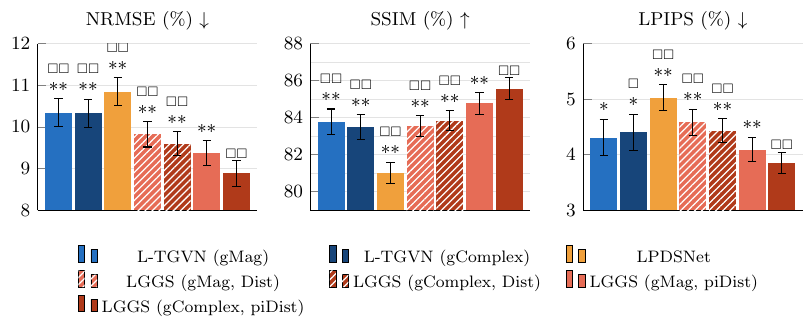}
    \caption{%
        Ablation over the guide representation (complex-valued vs.\ magnitude)
        crossed with the feature similarity (distance vs.\ the phase-invariant distance of Section~\ref{sec-phase-invariant}), at $8\times$ nominal acceleration.
        Two marker sets are stacked above each bar: $\ast$ tests against LGGS
        (gComplex, piDist), $\scriptstyle\square$ against LGGS (gMag, piDist).
    }
    \label{fig-results-bars-guidesim}
\end{figure}
\subsection{Guide Representation and Phase-Invariant Similarity}
\label{sec:res-abl-guidesim}
Figure~\ref{fig-results-bars-guidesim} shows quantitative results ablating the four guide representation
(complex or magnitude) and feature similarity (the distance of
\eqref{eq-similarity-gcdl}, or the phase-invariant distance of
Section~\ref{sec-phase-invariant}) combinations, at the $8\times$ nominal operating point. 
All four combinations improved on the unguided LPDSNet.

We observed that the similarity was the larger of the two effects. Phase
invariance yielded $0.45$\% NRMSE with a magnitude guide and $0.72$\% with a complex
one, whereas the complex guide yielded $0.22$\% under the plain distance and $0.49$\%
once the similarity is made invariant to phase. The two axes are therefore not
independent: a complex guide is of most benefit once its patch-constant phase can be
discounted, which is the function of \eqref{eq-pidistsim}. This was corroborated
by the variational-net comparison, in which L-TGVN was supplied the identical prior
examination in both forms and did not distinguish between them ($10.33$\% NRMSE
in either case).

Appendix~\ref{app:visuals} compares the four configurations
visually, on two cases. We observed that the guide visit had a substantially
different phase profile from the reconstruction, and that reconstructions
employing the phase-invariant similarity recovered much more image detail, which
was clearly available in the guide visits. Appendix~\ref{app:guide-rep} isolates
what a magnitude guide discards.

\begin{figure}[tbh]
    \centering
    \includegraphics[width=\linewidth]{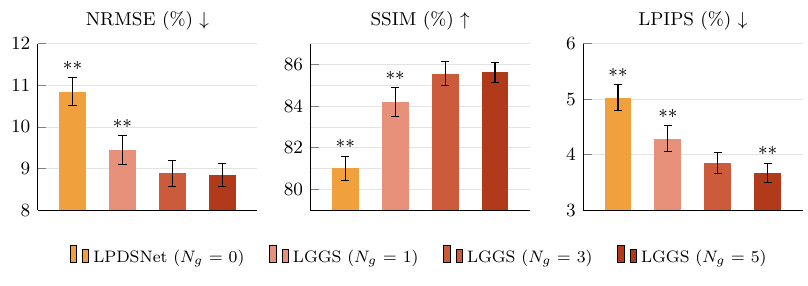}
    \caption{%
        Ablation over the number of guide slices $N_g$ at $8\times$ nominal acceleration.
        Markers test against LGGS ($N_g = 3$), the configuration used in every other figure.
    }
    \label{fig-results-bars-nguides}
\end{figure}

\begin{figure}[tbh]
    \centering
    \includegraphics[width=\columnwidth]{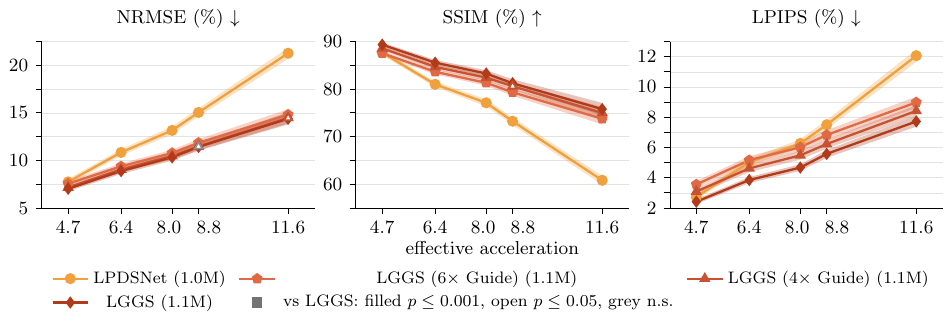}
    \caption{%
        Reconstruction quality when the guide is itself an accelerated
        reconstruction rather than a fully sampled prior
        (Section~\ref{sec:res-abl-guideaccel}).
    }
    \label{fig-results-accel-guides}
\end{figure}
\subsection{Number of Guide Slices}
\label{sec:res-abl-nguides}
Figure~\ref{fig-results-bars-nguides} shows the quantitative effect of varying the number of
guide slices $N_g$, with all else held fixed, at the $8\times$ nominal acceleration rate. We observed that a single guide
slice recovered the majority of the available gain, and that a three-slice window recovered a significant
further part of it, while including up to 5 guide slices moved neither NRMSE
nor SSIM measurably.

We note that the number of slices which may meaningfully aid reconstruction
is determined by the severity of inter-visit misalignment. A wide window
remains safe, as a guide slice which does not match is down-weighted rather than
averaged in. Widening $N_g$ beyond the useful width incurs
computational cost rather than a loss in accuracy. 
Representative visual examples are given in Appendix~\ref{app:visuals}.

\subsection{Accelerated Guide Images}
\label{sec:res-abl-guideaccel}
The preceding experiments supplied a fully sampled prior examination, which may
not be available in deployment. In this section we replaced it with a
reconstruction produced once by a pretrained LPDSNet~\cite{lpds2025} at
$4\times$ or $6\times$, held fixed while the target acceleration rate was swept.

Figure~\ref{fig-results-accel-guides} gives quantitative results for using
reconstruction-based guide images, showing that the resulting degradation was
small. At $16\times$ nominal acceleration, a $4\times$ guide cost $0.09$\% NRMSE against the fully sampled prior and a $6\times$ guide cost $0.44$\%, whereas the
absence of any guide cost a $6.89$\% penalty. 
At $12\times$ nominal acceleration, the $4\times$
guide was not significantly different from the fully sampled one in NRMSE. 
These results show that a prior examination need not be fully sampled in order to be
informative, making the method applicable to an archive of accelerated
studies. A visual comparison is given in Appendix~\ref{app:visuals}.

\section{Discussion and Conclusion}\label{sec-discussion}
In this work, we developed LGGS by adapting the learned group-sparsity
formulation of GroupCDL~\cite{janjusevic2025groupcdl} to longitudinal MRI
reconstruction within the analysis-sparsity framework of
LPDSNet~\cite{lpds2025}. The proposed method incorporates patient-specific prior
information directly into the reconstruction regularizer through explicit
feature correspondence across exams. Across different acceleration rates, LGGS
achieved superior reconstruction quality compared with two unguided
reconstruction methods and L-TGVN while requiring substantially fewer learned
parameters. These results demonstrate the potential of guided group sparsity to
exploit prior patient information for highly accelerated MRI reconstruction.

A major innovation of LGGS is the explicit incorporation of longitudinal
information through guided group sparsity. Conventional sparse regularization
can suppress weak image features at high acceleration rates, even when these
features represent true anatomical structures. LGGS addresses this limitation by
using corresponding features from prior exams to adapt the regularization
applied to the current reconstruction. This allows features with strong
longitudinal correspondence to receive less suppression and potentially
preserves anatomical details that would otherwise be lost. Unlike the original
GroupCDL, which exploits non-local similarities within a single image, LGGS
focuses its non-local group associations on features across exams. These prior
features provide more relevant patient-specific anatomical information that is
not available in the current undersampled data. The substantial improvement
over the unguided LPDSNet, particularly at higher acceleration rates,
demonstrates the value of longitudinal information beyond the general image
priors learned during supervised reconstruction.

Another important innovation is the explicit identification of corresponding
features between the current and prior exams through circulant attention. LGGS
evaluates feature similarity within a local search window rather than assuming
exact spatial correspondence between exams. This allows the method to
accommodate local spatial misregistration after global alignment. Multiple
neighboring guide slices further improve robustness to through-plane
misregistration, as demonstrated by the improvement with three or five guide
slices compared with a single slice (Section~\ref{sec:res-abl-nguides}).
Importantly, the similarity-based weights also reduce the contribution of prior
features that do not match the current reconstruction, which is particularly
relevant when anatomical or pathological changes occur.

The proposed phase-invariant similarity further addresses an important challenge
associated with complex-valued longitudinal MRI. Even when the underlying
anatomy is unchanged, MRI phase may vary substantially across exams due to
variations in magnetic field inhomogeneity, coil loading, and coil
sensitivity estimation~\cite{robinson2017phase}. Conventional complex-valued similarity
measures may therefore fail to identify corresponding anatomical structures.
LGGS explicitly accounts for these phase differences while preserving useful
relative complex feature information. Unlike magnitude-only similarity, which
discards all phase information, the proposed method removes only the common
phase difference between compared feature vectors. Our ablation experiments
demonstrate that phase-invariant similarity improves reconstruction quality and
that complex-valued prior images offer additional information beyond
magnitude-only guidance (Section~\ref{sec:res-abl-guidesim}), with further
theoretical and visual justification given in Appendix~\ref{app:guide-rep}. 
These results highlight the importance of preserving complex-valued information
from prior exams and developing appropriate methods to exploit it effectively.

LGGS differs from other longitudinal reconstruction methods in how prior
information influences reconstruction. L-TGVN's trust projection~\cite{ltgvn2026} usefully reduces interference with the acquired
measurements, but the correspondence between prior and current anatomy is
handled implicitly by its guidance network. LGGS instead evaluates local feature
correspondence explicitly, and uses the prior only to adapt the learned
group-thresholding operation, which prevents the direct transfer of prior-image
feature values.

LGGS is also compact. At $1.08$M learned parameters it reconstructs
better than L-TGVN ($43.6$M) at every rate evaluated, and at comparable wall-clock cost
(Table~\ref{tab:bench}). Explicit incorporation of longitudinal information
through guided group sparsity is therefore an efficient alternative to a
substantially larger learned guidance network, with correspondingly lower
storage and compute demands. As Section~\ref{sec:setup} notes, the comparison
holds under the stated experimental conditions rather than against L-TGVN's
full-scale form.

We further demonstrate that LGGS does not require a fully sampled prior exam to
provide effective guidance. Prior images reconstructed from accelerated
acquisitions at $4\times$ or $6\times$ continued to provide substantial
improvements over unguided reconstruction, with only modest degradation relative
to fully sampled guide images (Section~\ref{sec:res-abl-guideaccel}). This is
particularly relevant for clinical applications, where prior MRI exams may
already have been acquired with accelerated protocols. Therefore, the ability to
use these prior accelerated images in LGGS further improves the practical value
of longitudinally informed reconstruction.

Several limitations should be acknowledged. First, the present study evaluated
only T1-weighted brain MRI, where inter-visit anatomical deformation is
generally less pronounced than in many body organs. Although LGGS accommodates
residual translation and through-plane misregistration, the current
implementation does not explicitly address large rotations or large anatomical
deformation. Second, the experiments used retrospective undersampling, and
further validation with prospectively accelerated acquisitions is warranted.
Although the clinical dataset includes patients with brain tumors and the
qualitative results demonstrate preservation of anatomical details, a systematic
evaluation of lesion preservation and true interval pathological changes would
provide additional evidence of clinical robustness. Future studies will also
explore longitudinal reconstruction in other anatomies, multi-contrast guidance,
and the use of multiple prior exams.

In conclusion, LGGS exploits patient-specific prior information through guided
group sparsity, explicit feature correspondence, and phase-invariant similarity,
improving reconstruction quality at high acceleration rates. Its compact
architecture and tolerance of accelerated prior images suit it to serial MRI
exams in practice.

\bibliographystyle{IEEEtran}
\bibliography{IEEEabrv,references}

\section*{Appendix}
\setcounter{figure}{0}
\renewcommand{\thefigure}{S\arabic{figure}}
\setcounter{table}{0}
\renewcommand{\thetable}{S\arabic{table}}
\setcounter{subsection}{0}
\renewcommand{\thesubsection}{\Alph{subsection}}

\subsection{Longitudinal Data Preparation}
\label{app:data-prep}

This section gives the preparation pipeline summarized in
Section~\ref{sec:setup}. Each subject's volumes are brought to a common axial
frame, trimmed to a shared slice range, and registered to visit $1$; the three
steps which bear on the reconstruction itself follow.

\noindent\textbf{Low-Frequency Translation Registration}:
Inter-visit alignment uses only the low-frequency content of each volume, which
makes the displacement estimate insensitive to contrast change and pathology.
For visit $t$ we form the band-limited coil-combined reference
${\longi{\rho}{t} = \mathrm{rss}(\cwise{\bF^\herm \bW_{\mathrm{ACS}}
\bF}\,\longi{\x}{t})}$, where $\longi{\x}{t} \in \C^{NC}$ is the multi-coil
volume, $\bW_{\mathrm{ACS}}$ Hamming-windows k-space and retains only the
central ACS band, and $\mathrm{rss}$ denotes root-sum-of-squares
coil combination. Visits $t = 2,3$ are registered to visit $1$ by the
integer-voxel 3D translation ${\bt_t = \argmax_{\bt \in \mathbb{Z}^3}
\langle \longi{\rho}{1},\, \mathcal{T}_{\bt}\longi{\rho}{t}\rangle}$, so that
in-plane and through-plane misregistration are estimated together. The shift is
applied alike to the multi-coil volume, the sensitivity maps and the support
mask. Rotation is not corrected, and is left to future work.

\noindent\textbf{Sensitivity Maps and Brain Mask}:
The prepared volumes carry Walsh~\cite{walsh2000adaptive} maps estimated from
the full multi-coil volume, and a per-slice brain-support mask
taken from
ESPIRiT~\cite{espirit} maps computed on the fully sampled
volume. Neither set enters the forward operator \eqref{eq-forward}, which is
built instead from the ACS band of the \emph{undersampled} k-space
(Section~\ref{sec:setup}), so that training and inference see maps of identical
k-space support. The mask alone is retained, and is used to zero the background
of the reconstruction target, rendering the training loss effectively in-brain,
and to window the display range in qualitative figures.

\noindent\textbf{ACS-Based Dynamic-Range Normalization}:
Coil-combined magnitudes vary substantially across subjects, scanners and
visits. Each visit is therefore divided by $2\longi{\alpha}{t}$, where
$\longi{\alpha}{t}$ is the median of $\longi{\rho}{t}$ over the foreground
voxels $\bM_t$. Taking the median over foreground only makes it insensitive to
background noise and to how much of the field of view the head occupies, and
reading it from the ACS band makes the same statistic available at inference.

\subsection{On Mean-Subtraction}
\label{app:mean-subtraction}
This section derives the geometry-aware mean-subtraction applied to the
LPDSNet/LGGS input.
Suppose we want to recover the mean-subtracted image ${\tilde{\x} = \x - \mu}$.
We approximate $\mu$ from our data by taking the mean of the zero-filled data
(or extracting the value at the center of k-space), ${\mu = \mathrm{mean}(\ObsOp^\herm \y)}$.

In our unrolling objective's data-fidelity, from which we derive part of our architecture,
we may consider the following change of variable,
\begin{align}
    & \nabla_{\tilde{\x}}\tfrac{1}{2}\norm{\y - \ObsOp(\x - \mu + \mu)}_2^2 \\
    =& \nabla_{\tilde{\x}} \tfrac{1}{2}\norm{\y - \ObsOp\tilde{\x} - \ObsOp(\mu\bOne)}_2^2 \\
    =& -\ObsOp^\herm (\y - \ObsOp\tilde{\x} - \ObsOp(\mu\bOne)) \\
    =& \ObsOp^\herm\ObsOp\tilde{\x} - (\ObsOp^\herm\y - \ObsOp^\herm\ObsOp(\mu\bOne)) \\
    =& \ObsOp^\herm\ObsOp\tilde{\x} - \tilde{\y}.
\end{align}
It is for this reason that our input to LPDSNet/LGGS is the ``geometry-aware"
mean-subtracted zero-filled image, ${\tilde{\y} = (\ObsOp^\herm\y -
\ObsOp^\herm\ObsOp(\mu\bOne))}$, rather than the na\"{i}ve $\ObsOp^\herm\y - \mu$ (used by the original LPDSNet \cite{lpds2025})
which does not take into account coil sensitivity geometry.

The guide is treated separately. It enters no data-fidelity term, so no
geometry-aware correction applies, and the network instead receives
${\tilde{\w} = \w - \mu_{\w}}$, with ${\mu_{\w} = \mathrm{mean}(\w)}$ the
guide's own spatial mean, taken jointly over the $N_g$ guide slices. A magnitude
guide is centered in the same way, by $\mathrm{mean}(\abs{\w})$, so that both
representations reach the analysis operator with zero mean and their codes
occupy the same domain as the reconstruction's. The constants removed
necessarily differ, as
${\abs{\mathrm{mean}(\w)} \neq \mathrm{mean}(\abs{\w})}$ for any image
carrying phase variation.

\subsection{Derivation of Phase-Invariant Similarity and Backpropagation Rules}
\label{app:phase-invariant}

This section provides the derivations supporting the phase-invariant
similarity functions introduced in
Section~\ref{sec-phase-invariant} of the main text.  We derive (i)~the
closed-form solutions of the phase-optimization problem
\eqref{eq-phase-invariant-def} for both the dot-product and
half-squared-distance similarity functions, and (ii)~the back-propagation
rules required to train the LGGS network end-to-end through these similarity
operators.

\subsubsection{Motivation: Failure Modes of Na\"{i}ve Similarities}

A magnetic resonance image $\x \in \C^N$ is inherently complex-valued, with a
corresponding magnitude image ${\br = \abs{\x}}$ and phase image
${\bvarphi = \angle\x}$.  The non-local similarity prior employed in LGGS
encourages latent vectors of similar non-local image regions to share the
same representation. Visually, similar image regions correspond to similar
patches of the magnitude image $\br$.  However, the image's phase may vary
throughout the image---and between longitudinal acquisitions---irrespective
of edge or texture information.

For example, consider image patches ${\bp,\bq \in \C^{P^2}}$ that represent
similar magnitude content but differ in phase, ${\bq = e^{j\varphi_0}\bp}$, with
corresponding analysis code vectors ${\bu, \bv \in \C^{M}}$ such that
${\bv = e^{j\varphi_0}\bu}$. If we na\"{i}vely compute similarity between these
complex code-vectors by treating them as $2M$-channel real vectors, i.e.\ via
the real inner-product similarity,
\begin{align}
    \simfun( \bu, \bv )
    &= \RE\{ \bv \}^\top \RE\{ \bu \} + \IM\{ \bv \}^\top \IM\{ \bu \} \\
    &= \RE\{ \bv^\herm \bu \} \\
    &= \RE\{ e^{-j\varphi_0} \bu^\herm\bu \} \\
    &= \cos(\varphi_0) \norm{\bu}_2^2,
\end{align}
then we risk failing to recognize similar image regions when the phase
difference is close to $\pi/2$ (and inverting the sign of similarity for
phase differences near $\pi$).  In light of this, we seek a differentiable
phase-invariant latent similarity metric.

A first candidate is the inner product of code magnitudes,
${\simfun(\bu,\bv) = \abs{\bv}^\top\abs{\bu}}$, which trivially eliminates
phase but extraneously recognizes similarities. As a toy example, consider
the vectors ${\bu = [1; -1]}$ and ${\bv = [1;1]}$, representing an edge and a
constant image region, respectively.  The standard and phase-invariant dot-product similarities
correctly identify these as orthogonal (${\bv^\top\bu = \abs{\bv^\top\bu} = 0}$). However, the
dot-product of code-magnitudes erroneously recognizes these codes as just as
similar as they are to themselves (${\abs{\bv}^\top\abs{\bu} = \bv^\top\bv = \bu^\top\bu = 2}$).

\subsubsection{Closed-Form Phase-Invariant Similarities}

The phase-invariant similarity \eqref{eq-phase-invariant-def} is defined by
phase-aligning the code vectors prior to similarity computation,
\begin{align}
    \pisim( \bu, \bv ) &= \max_\varphi ~ \simfun(\bu, \, e^{j\varphi} \bv).
\end{align}
For the real inner product, the phase-invariant dot-product similarity becomes,
\begin{align}
    \pidotsim( \bu, \bv ) &= \max_\varphi ~ \RE\{ e^{-j\varphi} \bv^\herm\bu \} \\
    &= \max_\varphi ~ \RE\{ e^{-j\varphi} e^{-j\varphi_0} \} \abs{\bv^\herm\bu} \\
    &= \abs{\bv^\herm\bu},
\end{align}
where ${\bv^\herm\bu = \abs{\bv^\herm\bu}e^{-j\varphi_0}}$ and the maximum is
achieved at $\varphi = -\varphi_0$.  Hence, the phase-optimal dot-product
similarity is simply the magnitude of the complex dot-product between the
two vectors, ${\pidotsim(\bu,\bv) = \abs{\bv^\herm\bu}}$.

In a similar manner, we express the phase-optimal distance-similarity as,
\begin{align}
    \pidistsim( \bu, \bv )
    &= \max_\varphi ~ -\tfrac{1}{2}\norm{\bu - e^{j\varphi}\bv}_2^2 \\
    &= \max_\varphi ~ -\tfrac{1}{2}\!\left(\norm{\bu}_2^2 - 2\RE\{ e^{-j\varphi}\bv^\herm\bu \} + \norm{\bv}_2^2\right) \\
    &= -\tfrac{1}{2}\!\left(\norm{\bu}_2^2 - 2\abs{\bv^\herm\bu} + \norm{\bv}_2^2\right),
\end{align}
recovering~\eqref{eq-pidistsim}.

\subsubsection{Back-Propagation Rules}

To enable end-to-end training through $\pidotsim$ and $\pidistsim$, we
derive their back-propagation rules.  We first write the phase-invariant
dot-product similarity of the latent matrices ${\X, \Y \in \C^{Q\times M}}$
as ${\SIMMAT = f(g(\X, \Y))}$ where ${f(\bZ)_{ij} = \abs{\bZ_{ij}}}$ and
${g(\X,\Y) = \X\Y^\herm}$.

For complex scalar $z = a + jb$, we derive the back-prop rule for $s=f(z)$
as follows,
\begin{align}
    s & = \abs{z} \\
    \dot{s} &= \frac{a\dot{a} + b\dot{b}}{\abs{z}}
        = \RE\left\{ \frac{ z^\ast \dot{z} }{ \abs{z} } \right\} = \RE\{ \sign(z)^\ast \dot{z} \} \\
    \RE\{ \dd{s}^\ast \dot{s} \} &= \RE\left\{ (\dd{s} \sign(z))^\ast \dot{z} \right\} \\
    \RE\{ \dd{s}^\ast \dot{s} \} &= \RE\left\{ \dd{z}^\ast \dot{z} \right\}
\end{align}
Hence, ${\dd{z} = \sign(z) \dd{s}}$.

Next we derive the back-prop rule for $\bZ = g(\X, \Y)$,
\begin{align}
    \bZ &= \X\Y^\herm \\
    \dot{\bZ} &= \dot{\X}\Y^\herm + \X\dot{\Y}^\herm \\
    \RE\ip{ \dd{\bZ} }{ \dot{\bZ} } &=
        \RE\ip{ \dd{\bZ} }{ \dot{\X}\Y^\herm } + \RE\ip{ \dd{\bZ} }{ \X\dot{\Y}^\herm } \\
    \RE\ip{ \dd{\bZ} }{ \dot{\bZ} } &=
        \RE\ip{ \dd{\bZ}\Y }{ \dot{\X} } + \RE\ip{ \X^\herm\dd{\bZ} }{ \dot{\Y}^\herm } \\
    \RE\ip{ \dd{\bZ} }{ \dot{\bZ} } &=
        \RE\ip{ \dd{\X} }{ \dot{\X} } + \RE\ip{ \dd{\Y}^\herm }{ \dot{\Y}^\herm }
\end{align}
Hence, ${\dd{\X} = \dd{\bZ}\Y}$ and ${\dd{\Y} = \dd{\bZ}^\herm\X}$.

Together, the back-prop rules for $\SIMMAT = f(g(\X, \Y))$ are,
\begin{gather}
    \dd{\X} = \dd{\bZ}\Y
        ~~\text{ and }~~
    \dd{\Y} = \dd{\bZ}^\herm \X, \\
        \text{where }
    \dd{\bZ} = \sign(\X\Y^\herm) \circ \dd{\SIMMAT}.
\end{gather}

For the phase-invariant distance-similarity $\pidistsim$, the back-propagation
rules follow from the same identities together with derivatives of the
quadratic norm terms $\norm{\X_{i:}}_2^2$ and $\norm{\Y_{i:}}_2^2$, yielding,
\begin{gather}
    \dd{\X} = \dd{\bZ}\Y - (\dd{\SIMMAT}\bOne) \circ \X  \\
        ~~\text{ and }~~ \notag \\
    \dd{\Y} = \dd{\bZ}^\herm\X - (\dd{\SIMMAT}^\top \bOne) \circ \Y, \\
        \text{where }
    \dd{\bZ} = \sign(\X\Y^\herm) \circ \dd{\SIMMAT}.
\end{gather}
See the supplementary material of \cite{janjusevic2025groupcdl} for further
details.

\subsection{Supplementary Visual Comparisons}
\label{app:visuals}

\begin{figure*}[tb]
    \centering
    \includegraphics[width=\linewidth]{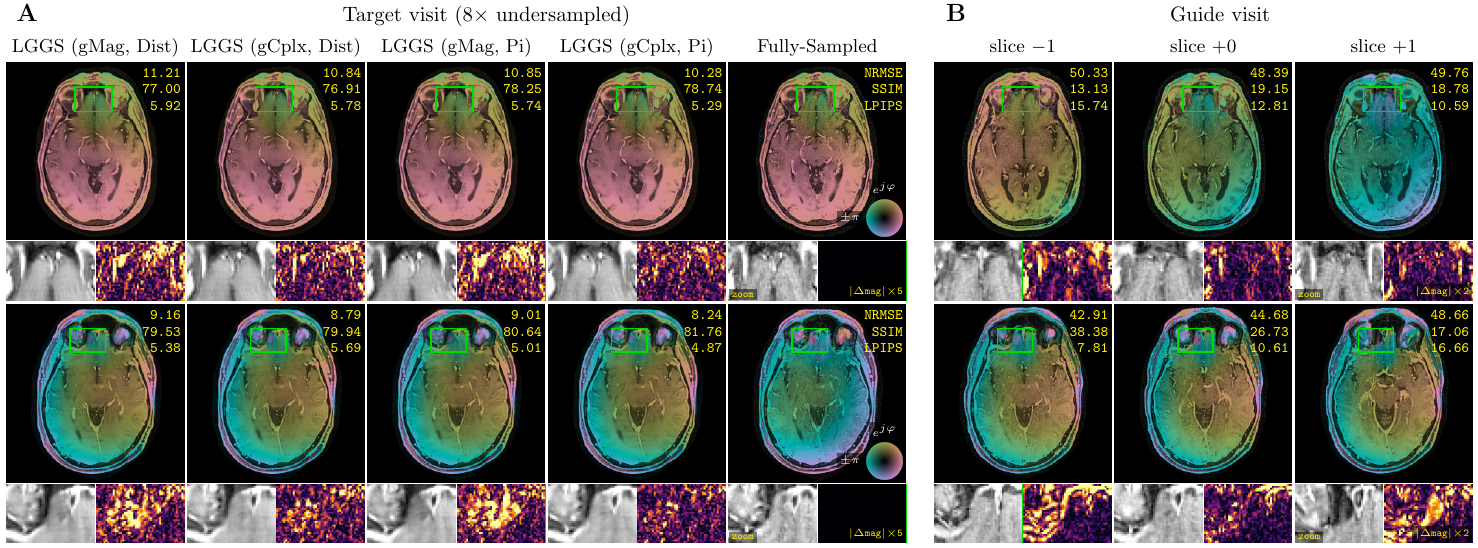}
    \caption{%
        The four guide (complex-valued vs.\ magnitude) and similarity (distance
        vs.\ the phase-invariant distance) combinations at $8\times$ nominal
        acceleration, on two cases. Panels are complex-valued with hue encoding
        phase; the magnified insets are magnitude.
        NRMSE\,/\,SSIM\,/\,LPIPS ($\times 100$) are given above each panel.
    }
    \label{fig-results-compare-guidesim}
\end{figure*}

\begin{figure}[tbh]
    \centering
    \includegraphics[width=\columnwidth]{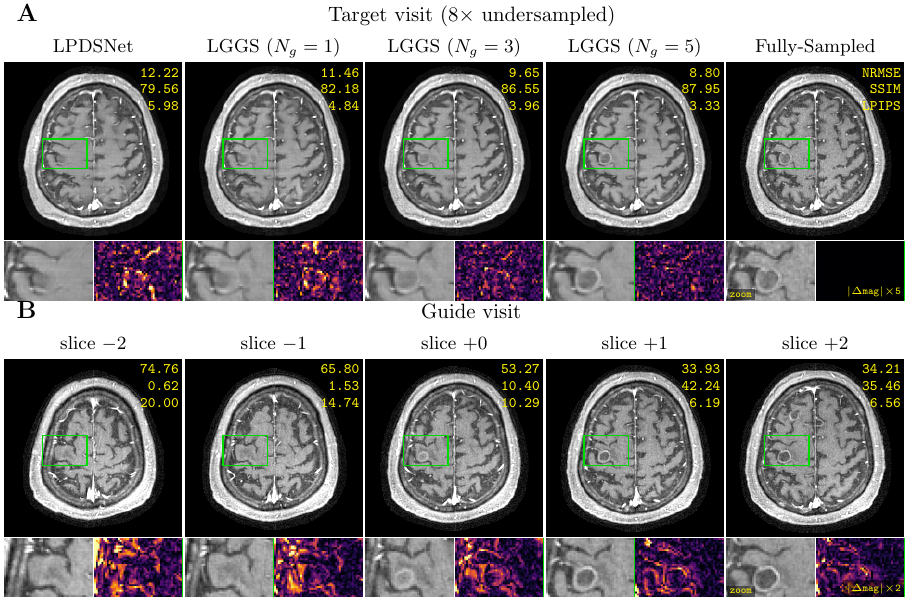}
    \caption{%
        Guide-slice ablation for a case of severe through-plane misregistration at
        $8\times$. \textbf{(A)} reconstructions, unguided then
        $N_g = 1,\,3,\,5$; \textbf{(B)} the guide slices those runs were given,
        the window centered on the co-located slice. Here slice $+1$ matches the
        target better than slice $+0$ ($33.93$\% against $53.27$\% NRMSE), so
        widening the window keeps paying, from $9.65$\% at $N_g=3$ to $8.80$\% at
        $N_g=5$. Figure~\ref{fig-app-nguides-mild} gives a mild case, in which
        $N_g=3$ suffices.
    }
    \label{fig-app-nguides-severe}
\end{figure}

\begin{figure}[tb]
    \centering
    \includegraphics[width=\columnwidth]{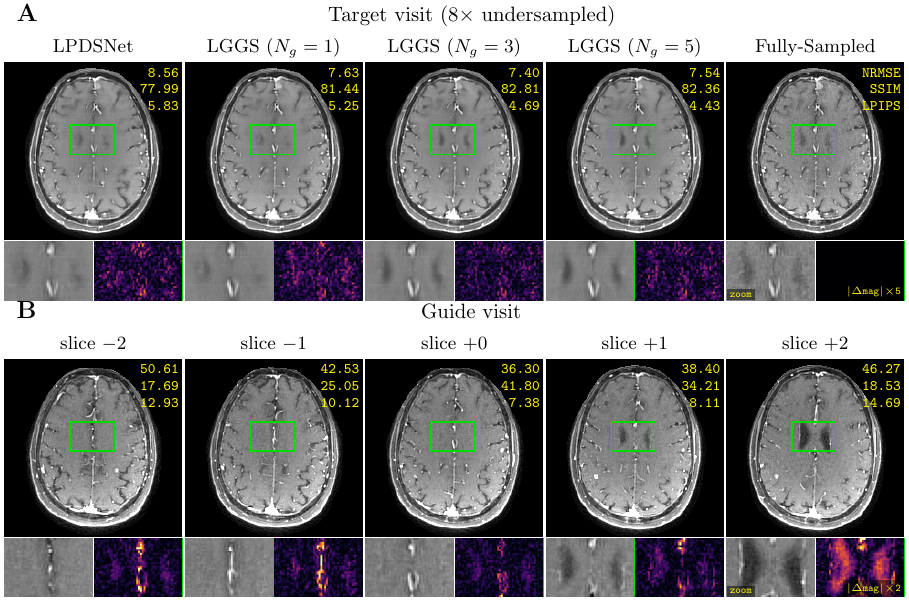}
    \caption{%
        Visual ablation over the number of guide slices, for a case of
        \emph{mild} through-plane misregistration at $8\times$ acceleration;
        the companion to Figure~\ref{fig-app-nguides-severe}, which gives a
        severe case. Here slice $+0$ is already the best match at $36.30$\%
        NRMSE against the target, the three-slice window captures what is
        available, and $N_g=5$ adds nothing on NRMSE or SSIM
        ($7.54$\%\,/\,$82.36$\% against $7.40$\%\,/\,$82.81$\%), improving only LPIPS.
    }
    \label{fig-app-nguides-mild}
\end{figure}

\begin{figure*}[tb]
    \centering
    \includegraphics[width=\linewidth]{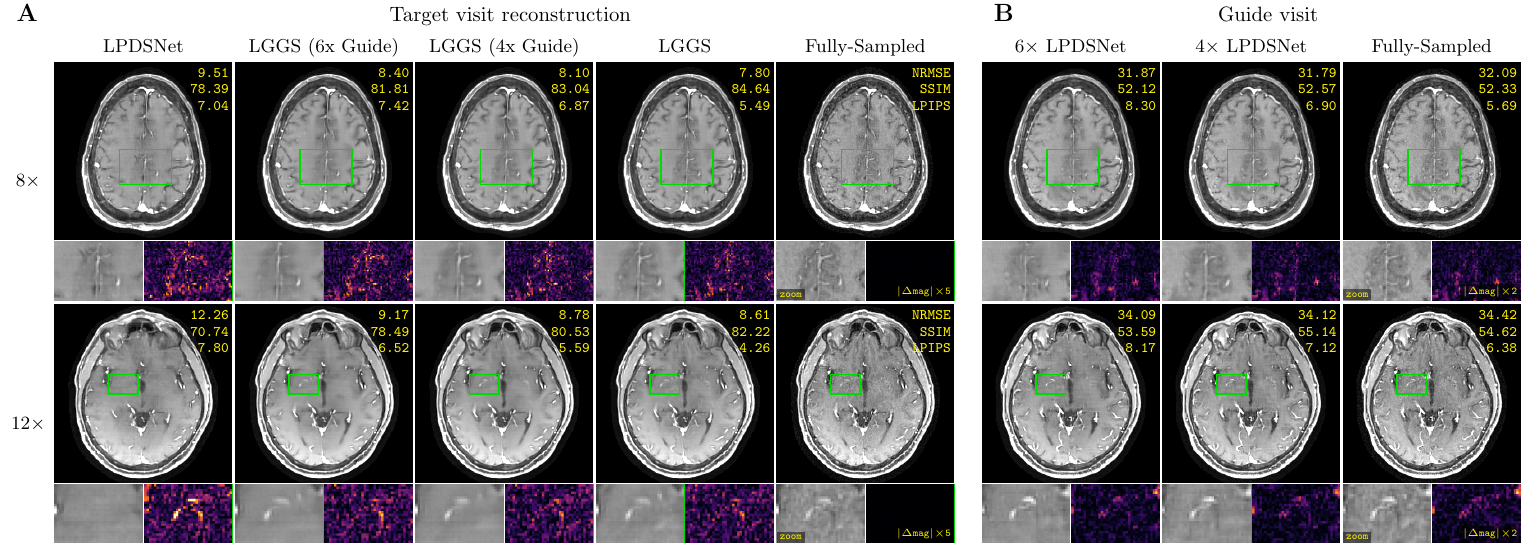}
    \caption{%
        Visual comparison with an accelerated guide, supporting
        Section~\ref{sec:res-abl-guideaccel}. \textbf{(A)} Unguided
        reconstruction, LGGS given the accelerated guide, LGGS given the
        fully sampled guide, and the target. \textbf{(B)} The guide as supplied
        to the network alongside the fully sampled prior it approximates, which
        bounds how much of the degradation is attributable to the guide itself.
        NRMSE\,/\,SSIM\,/\,LPIPS ($\times 100$), with respect to the
        fully sampled target, are given above each panel.
    }
    \label{fig-app-guideaccel}
\end{figure*}

This section collects the visual comparisons deferred from
Section~\ref{sec:results}.
Figure~\ref{fig-results-compare-guidesim} gives the guide\,$\times$\,similarity
ablation on two cases, Figures~\ref{fig-app-nguides-severe}
and~\ref{fig-app-nguides-mild} give the guide-slice ablation on a severe and a
mild case of through-plane misregistration, and Figure~\ref{fig-app-guideaccel}
compares reconstructions from an accelerated against a fully sampled guide.

\subsection{Complex and Magnitude Guides}
\label{app:guide-rep}

Section~\ref{sec:res-abl-guidesim} finds a complex guide preferable to a
magnitude one even once the similarity is made phase-invariant. The two are not
alternatives: the invariance removes far less phase than the magnitude does.

Phase-invariant similarity $\pidistsim$ maximizes over a single scalar rotation per query--key pair, so it
discards one phase per comparison. Fix a pixel $n$, write
${\w = \br \odot e^{j\bvarphi}}$, and split the phase on the support of the
analysis filter at $n$ into its mean (among the filter support) and the deviation about it,
${\bvarphi = \bar\varphi + \bdelta}$. The linearity of $\bA^\herm$ then
gives,
\begin{equation}\label{eq-guide-decomp}
    \bv^{\C} = e^{j\bar\varphi}\bigl(\bv^{\mathrm{M}} + \bb\bigr),
\end{equation}
where $\bv^{\C}$ and $\bv^{\mathrm{M}}$ are the codes of $\w$ and $\br$ at $n$,
and $\bb$ is the code there of ${\br \odot (e^{j\bdelta} - 1)}$. The leading
rotation is precisely what the invariance absorbs; what it cannot absorb is
$\bb$, which vanishes if and only if ${\bdelta \equiv 0}$. A magnitude guide
discards the phase at every pixel of the support, whereas $\pidistsim$ discards
one constant across it, and $\bb$ is the difference. 

\begin{figure}[tb]
    \centering
    \includegraphics[width=\columnwidth]{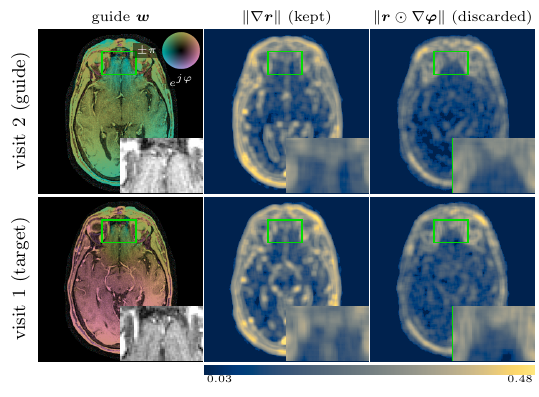}
    \caption{%
        What a magnitude guide discards, for the first case of
        Figure~\ref{fig-results-compare-guidesim}. The local variation of the
        fully sampled examination splits into $\norm{\nabla\br}$, which a
        magnitude guide retains, and $\norm{\br \odot \nabla\bvarphi}$, which
        it discards;
        both are pooled over the $7\times7$ analysis support, the scale at which
        the codes are formed, and share one color key, so their magnitudes may
        be read against each other. The discarded channel holds structure the
        retained one does not, and repeats between visits. Hue gives phase in
        the first column.
    }
    \label{fig-app-guide-phase}
\end{figure}

\begin{figure}[tb]
    \centering
    \includegraphics[width=\columnwidth]{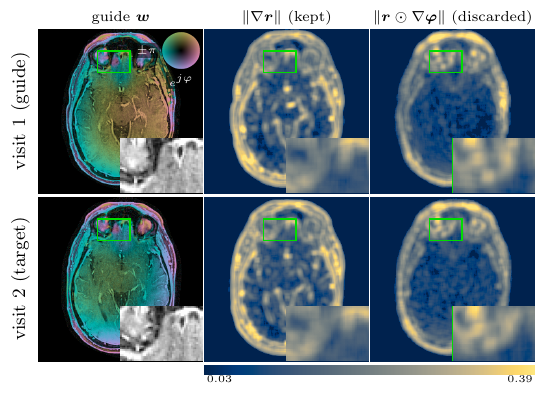}
    \caption{%
        As Figure~\ref{fig-app-guide-phase}, for the second case of
        Figure~\ref{fig-results-compare-guidesim}.
    }
    \label{fig-app-guide-phase2}
\end{figure}

Whether $\bdelta$ is worth keeping is then a question about the prior
examination rather than about the architecture. Writing the local variation of
the guide as
${\nabla\w = (\nabla\br + j\,\br \odot \nabla\bvarphi)\,e^{j\bvarphi}}$,
whose two channels are orthogonal in the $e^{j\bvarphi}$ frame, a magnitude
guide retains $\norm{\nabla\br}$ and discards
$\norm{\br \odot \nabla\bvarphi}$ entirely.
Figures~\ref{fig-app-guide-phase} and~\ref{fig-app-guide-phase2} show both
channels for the two cases of Figure~\ref{fig-results-compare-guidesim}. The
discarded channel carries the susceptibility-driven contrast of the prior
examination~\cite{robinson2017phase}, marking vessels and tissue boundaries that
the magnitude image need not resolve, and it is anatomy rather than acquisition
noise: between two visits of one patient it correlates at $0.73$, against $0.23$
between two different patients.

\subsection{Masked and Unmasked SSIM}
\label{app:unmasked}

\begin{figure*}[tb]
    \centering
    \includegraphics[width=\linewidth]{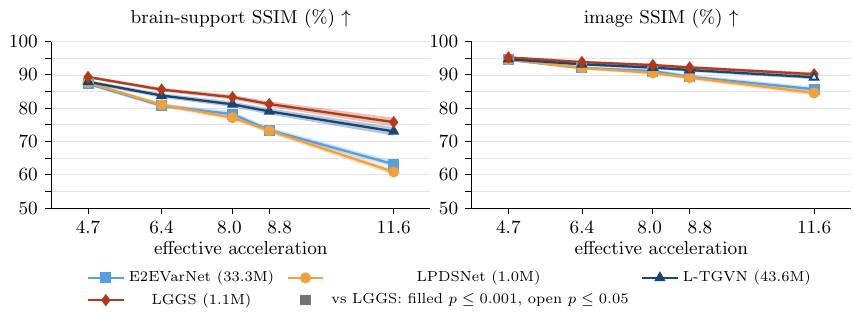}
    \caption{%
        SSIM against effective acceleration, averaged over the brain-support
        mask (left) and over the whole frame (right), on a shared vertical
        scale. Otherwise as the SSIM panel of
        Figure~\ref{fig-results-accel}.
    }
    \label{fig-app-unmasked}
\end{figure*}

Figure~\ref{fig-app-unmasked} sets the SSIM of Figure~\ref{fig-results-accel},
averaged over the brain-support mask, beside the same quantity averaged over the
whole frame. The mask covers $32\%$ of the frame, and the background, where
target and reconstruction are both zero, scores above $99$\% throughout.
Averaging over the frame raises every curve and flattens them together: between $4.7$
and $11.6\times$ LGGS falls by $13.5$\% on the mask but only $5.0$\% over the
frame, and at $11.6\times$ the spread across the four methods narrows from
$15.0$\% to $5.6$\%. The ordering survives, but a reader comparing methods on the
unmasked number sees a third of the difference.

\subsection{Architecture Hyperparameters}
\label{app:arch}

Tables~\ref{tab:arch-unrolled} and~\ref{tab:arch-blackbox} give the
hyperparameters of the two architecture families compared in
Section~\ref{sec:setup}. Parameter counts are real-valued throughout: a complex
weight counts as two parameters and a weight tied across layers is counted
once, placing the complex-valued unrolled networks and the real-valued
variational-net based architectures on a common footing.

\begin{table}[tb]
\centering
\caption{Unrolled sparse-coding architectures. $p$ is the filter support and
$W_z$/$W_g$ the self- and guide-similarity windows. LPDSNet forms no adjacency,
so its attention columns do not apply.}
\label{tab:arch-unrolled}
\vspace{0.3em}
\resizebox{\linewidth}{!}{%
\begin{tabular}{l|cccccccc} \hline
    Name & $K$ & $p$ & $M$ & $M_h$ & $W_z$ & $W_g$ & $N_g$ & params \\ \hline
    LPDSNet & 30 & 7 & 169 & -- & -- & -- & 0 & 1.00\,M \\
    LGGS    & 30 & 7 & 169 & 64 & 1  & 3  & 3 & 1.08\,M \\ \hline
\end{tabular}}
\end{table}

\begin{table}[tb]
\centering
\caption{Variational-net architectures. L-TGVN adds the trust-region
projection $\mathcal{P}_\delta$ of \eqref{eq-tgvn}, solved by conjugate
gradient (CG).}
\label{tab:arch-blackbox}
\vspace{0.3em}
\resizebox{\linewidth}{!}{%
\begin{tabular}{l|cccccccc} \hline
    Name & cascades & \makecell{U-Net\\chan.} & pools &
    \makecell{sense\\chan.} & \makecell{sense\\pools} &
    \makecell{CG\\iters} & $N_g$ & params \\ \hline
    E2EVarNet & 12 & 18 & 4 & 8 & 4 & -- & 0 & 33.3\,M \\
    L-TGVN    &  6 & 18 & 4 & 8 & 4 & 10 & 3 & 43.6\,M \\ \hline
\end{tabular}}
\end{table}

\end{document}